\documentclass[aps,prb,onecolumn,longbibliography]{revtex4-2}
\pdfoutput=1
\usepackage[caption=false]{subfig}

\usepackage{array}
\usepackage{slashed,bbold}
\usepackage{physics}
\usepackage{dsfont}

\usepackage{amsmath,amssymb,bm} 
\usepackage{graphicx}
\usepackage{soul}
\usepackage{mhchem}

\usepackage{xcolor}
\usepackage{soul}

\usepackage[papersize={8.5in,11in}]{geometry}

\usepackage{ colortbl}
\usepackage{color} 
\definecolor{darkblue}{rgb}{0.,0.,0.4}
\definecolor{darkred}{rgb}{0.5,0.,0.}
\definecolor{BlueViolet}{RGB}{138,43,226}
\definecolor{SkyBlue}{RGB}{30,144,255}
\definecolor{DarkGreen}{RGB}{0,100,0}
\usepackage[pdftex,colorlinks=true,linkcolor=darkblue,citecolor=blue,urlcolor=darkred]{hyperref}

\def \be{\begin{equation}}
\def \ee{\end{equation}}

\begin{document}


	\title{Spin-Flux Skyrmions: Anomalous Electron Dynamics and Spin-Hall Currents} 
	\author{Sandip Bera and Sajeev John }
	\affiliation{Department of Physics, University of Toronto, 60 St. George Street, Toronto, Ontario, Canada M5S 1A7	}

\begin{abstract}		
We introduce a topologically distinct skyrmion, termed a spin-flux skyrmion, which shares the same real-space magnetization profile as a conventional skyrmion but differs fundamentally in its underlying topological structure. This distinction originates from the path traced by its rotation matrices within the doubly connected SO(3) group manifold, leading to a nontrivial spinor phase of  $e^{i\pi}$ upon encircling the texture. Using an explicit SU(2) gauge field formalism, we derive the emergent magnetic field components generated by both conventional and spin-flux skyrmions. While conventional skyrmions exhibit a dominant  $\sigma_z$ component with weak dipolar $\sigma_x, \sigma_y$ contributions, spin-flux skyrmions possess an additional monopolar $\sigma_x$ component that yields a finite average emergent field  for a finite density of  skyrmions. This nontrivial component introduces a nontrivial term in the Hall conductivity, enabling a direct explanation of  experimental Hall resistivity anomalies that cannot be accounted for by conventional skyrmions alone. Moreover, we show that this additional term couples to the in-plane spin polarization of conduction electrons, providing a further tunable handle to control the transverse Hall response.
\end{abstract}			
\maketitle

\section{Introduction}

Magnetic skyrmions are nanoscale spin textures that exhibit a characteristic whirl-like magnetization pattern \cite{Skyrme19611, Skyrme19612, Skyrme1962}, first observed in non-centrosymmetric magnets \cite{Muhlbauer2009}, and later found in many ferromagnetic \cite{Yu2012,Fert2013, Nagaosa2013} and antiferromagnetic \cite{Juge2022,Juge2022} systems. A key property of skyrmions is their integer winding number, often referred to as the skyrmion number, which distinguishes them from other magnetic configurations \cite{Rajaraman_book, Nagaosa2013}. This number measures how many times the local spin configuration wraps around the unit sphere $S^2$, as described by a smooth spin texture $\vec{M}(\vec{r})$, over the two-dimensional plane. It is given by $N_{sk}= \frac{1}{4\pi} \int \vec{M}\cdot(\frac{\partial\vec{M}}{\partial x}\times  \frac{\partial\vec{M}}{\partial y})dxdy$.  A value of $N_{sk}=1$ corresponds to a single skyrmion, while larger absolute values represent multiple windings, and the sign indicates the sense of rotation  of the configuration. Both theoretical and experimental studies have shown that skyrmions can remain stable over a wide range of external magnetic fields, magnetic anisotropy strengths, and temperatures \cite{sandip2019,sandip2020,Bera2024,Romming2013,Herve2018,Yu2018,Yu2010,MoreauLuchaire2016}. 

\begin{figure}
	\begin{center}
		\subfloat[\label{so3u1}]{
			\includegraphics[scale=0.25]{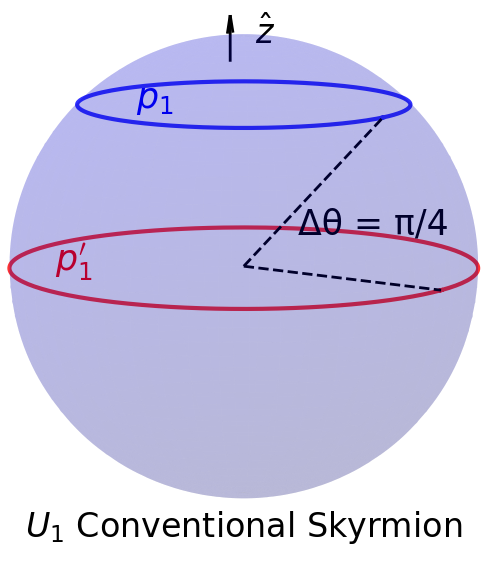}}~~
		\subfloat[\label{so3u2}]{				
			\includegraphics[scale=0.25]{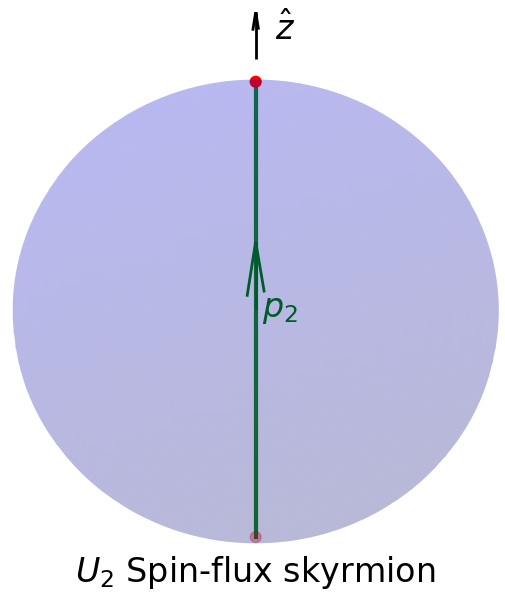}}
	\end{center}
	\caption{We illustrate the distinction between  two  skyrmions corresponding to  magnetic moment rotation fields $U_1$ and $U_2=U_2^zU_2^y$, in terms of the SO(3)  group manifold  (panels (a) and (b)). In this picture, a solid ball of radius $\pi$ is used to represent all possible spin rotations: each point inside the ball corresponds to a unique rotation, the displacement from the origin indicates the rotation angle, and the direction specifies the rotation axis. The surface of the ball therefore represents a rotation angle of $\pi$. For the conventional  $U_1$ skyrmion (panel (a)), we consider  circular paths in coordinates space with radii  $r=\rho_0$ (the skyrmion core radius) and $r=\infty$, which we label as $p_1$ and $p_1^{\prime}$, respectively. All rotation matrices describing this skyrmion configuration lie  infinitesimally belong the surface of the SO(3) ball (shown in  blue and red). Here, the rotation field corresponds to a fixed rotation angle of $\pi$ about the spatially varying axis unit vector $\hat{n}(\vec{r})$. As the skyrmion is encircled, the  spin rotation trajectory in SO(3) forms a closed loop on the surface.   Both paths, $p_1$ and $p_1^{\prime}$, can be continuously deformed to a single point and are considered topologically trivial within SO(3). For the $U_2$(spin-flux) skyrmion (panel (b)), the rotation $U_2^y$  corresponds to single point in SO(3) on the y-axis for any circular path of given radius. The rotation   $U_2^z$ traverses  a straight line path from the surface of SO(3), through the center of the ball, to the antipodal point as the skyrmion is encircled at any radius and $-\pi<\zeta<\pi$. This endows the $U_2$-skyrmion with a nontrivial spin-flux of $\pi$. Since antipodal points on the sphere of radius $\pi$  are  identical rotations, the path $p_2$  cannot be continuously deformed to a single point.  In other words, the  paths $p_1 (p_1^{\prime})$ and $p_2$ are homotopically distinct.}
	\label{so3_u1u2}
\end{figure}

The formation and stability of magnetic skyrmions results from a delicate competition between fundamental magnetic interactions \cite{Fert2017, Rohart2013}. The symmetric Heisenberg exchange interaction ($J$) favors parallel spin alignment. A crucial ingredient for inducing chirality is the Dzyaloshinskii-Moriya interaction (DMI) ($D$), an antisymmetric exchange term that favors a fixed, perpendicular canting between neighboring spins \cite{Dzyaloshinskii1957, Moriya1960}. This interaction requires broken inversion symmetry, which  occurs  in non-centrosymmetric crystal lattices. At low or zero external magnetic field, the competition between the DMI and Heisenberg exchange typically produces a one-dimensional spin-spiral (SS) ground state, where the magnetization rotates periodically with a wavelength proportional to $J/D$. Magnetic anisotropy,  such as an easy crystallographic axis for orientation, favors localized magnetic twists, rather than extended spirals.  Applying an external magnetic field, which favors spin alignment along the field direction ($+\hat{z}$),  likewise stabilizes localized magnetic textures relative to  the spin-spiral state. These textures satisfy a well-defined boundary condition: spins are aligned with the field ($+\hat{z}$) at large distances, while at their core, the spins point opposite to the field ($-\hat{z}$), maintaining a non-zero  integer winding number\cite{Camley2023}. As the magnetic field strength increases, the ground state evolves predictably. At low fields, the spin-spiral (SS) phase is stable. With increasing field, the system transitions into a skyrmion lattice phase. At very high fields, the system eventually reaches a fully polarized ferromagnetic (FM) state, in which all spins align with the external field. This sequence of phase transitions has been consistently observed in both theoretical models and experimental studies \cite{sandip2019,Bera2024,Romming2013,Herve2018,Yu2018,Yu2010,MoreauLuchaire2016}.

In spintronics, magnetic skyrmions are considered promising candidates for next-generation memory and logic devices due to their  stability and nanoscale dimensions \cite{MoreauLuchaire2016,Soumyanarayanan2017}, as well as their controllability  by electric currents. Their ability to carry information in a stable, well-defined spin configuration makes them ideal for high-density, low-power-consumption magnetic memory\cite{Schulz2012,Iwasaki2014}. Skyrmions can be driven by current densities two to four orders of magnitude lower than those required to move conventional magnetic domain walls, typically in the range of $10^6 - 10^8~\mathrm{A/m^2}$, compared to $10^{10} - 10^{12}~\mathrm{A/m^2}$ for domain wall motion \cite{Jonietz2010,Yu2012}. This substantial reduction in current for skyrmion motion, makes them attractive for energy-efficient spintronic devices. For example, skyrmion-based racetrack memory \cite{Park2014,Parkin2008,Fert2013} encodes information in the presence or absence of skyrmions along a narrow magnetic track. Data can be written, shifted, and read by applying short current pulses that move skyrmions, enabling faster data storage and retrieval. This architecture offers high storage density \cite{Beg2015,Woo2018} while consuming significantly less energy compared to conventional magnetic hard drives, which rely on moving mechanical parts or high current densities for domain wall motion. Moreover, the high sensitivity of skyrmions to  low driving current may enable  novel spintronic logic devices \cite{Zhang2015sci}.

Electron transverse deflection under different Lorentz-like forces leads to various Hall effects. The ordinary Hall effect arises from the Lorentz force of a physical magnetic field acting on charge carriers. In contrast, the anomalous Hall effect occurs in ferromagnetic materials with magnetic textures due to intrinsic spin-orbit coupling, even without an external magnetic field \cite{PhysRevLett.117.046601}. In materials hosting skyrmions, the spatial variation of magnetization gives rise to a synthetic  matrix magnetic field, strongest near the skyrmion cores. This field deflects conduction electrons, producing a transverse voltage. The resulting contribution to the Hall signal is referred to as the \emph{skyrmion-induced Hall effect} \cite{Fert2013, Nagaosa2013}.

In this article, we introduce a topologically distinct skyrmion that superficially exhibits the same magnetization profile as the conventional Neel skyrmion. The topological distinction arises from the detailed set of rotation matrices used to form the magnetic texture and the path traced by  the rotation matrices, within the SO(3) group manifold of physical rotations in three dimensional space, as the skyrmion is encircled by a path in  two-dimensional coordinate space  (Fig.\ref{so3_u1u2}). It is well known that the group manifold of SO(3) is topologically, doubly-connected. It has been suggested \cite{Nathan1951,Sajeev1995} that this is the fundamental reason for the existence of spin$-1/2$ particles. It enables the two-valued nature of the spin$-1/2$ wavefunction and that under a $2\pi$ rotation, the wavefunction changes sign. We refer to  this distinctive magnetic soliton as a spin-flux skyrmion. As in the case of conventional skyrmions, the effects of the spin-flux skyrmion on a conduction electron can be described by a matrix gauge field whose matrix components can be represented by the set of Pauli spin matrices $\{\sigma_x, \sigma_y, \sigma_z\}$. A fundamental consequence of the spin-flux skyrmion is that when an electron electron encircles this texture it acquires a nontrivial phase of $e^{i\pi}$. It has been suggested that nontrivial and consequential spin-flux  may play a  vital role in high-temperature cuprate superconductors  \cite{Sajeev1995,PhysRevB.69.224515,PhysRevB.61.16454,PhysRevB.51.12989,S.John1998}

In this paper, we focus on the explicit form of the SU(2) gauge field for spin-flux skyrmions and highlight its distinctive features compared to conventional skyrmions. For conventional skyrmions, the $\sigma_z$ component of the emergent magnetic field is dominant, peaking at the skyrmion core and rapidly decaying with increasing radius, while the $\sigma_x$ and $\sigma_y$ components exhibit a weaker dipolar distribution, with a peak magnitude approximately one quarter of that of $\sigma_z$. In contrast, spin-flux skyrmions exhibit a qualitatively different behavior: the $\sigma_z$ component retains a similar radial profile, but  a highly consequential $\sigma_x$ component appears,  with a singular monopolar character. This monopolar feature leads to a finite contribution to the average emergent field in a skyrmion lattice, a property absent in the conventional case. Consequently, the emergent magnetic  field introduces an additional term in the Hall conductivity, supplementing the diagonal contribution common to both skyrmion types. This additional term provides a natural explanation for certain experimental observations that cannot be  captured by the conventional skyrmion model alone. Furthermore, we demonstrate how the spin polarization density matrix of  electrons driven by an electric field can be used to control the Hall current in the presence of spin-flux skyrmions.

\section{Emergent   magnetic field governing  skyrmion-electron interaction}
 
\subsection{Topological Distinctions}
 
The interaction between skyrmions and electrons arises due to the coupling between the spin of conduction electrons and the local magnetization of the skyrmion texture. This interaction can be effectively described by an exchange coupling term in the Hamiltonian, where the conduction electron spin $\vec{\sigma}$ interacts with the local magnetization $\vec{M} = (\sin\theta\cos\phi, \sin\theta\sin\phi,\cos\theta) $ of the skyrmion via $H_{int}= -J_{h} \vec{M}\cdot  \vec{\sigma}$, and   $J_{h}$ is the exchange interaction strength \cite{Zhang2009,Hamamoto2015,Ohgushi2000,Tome2021,Osca2020,Han:2017fyd}. The orientation angles $\theta$ and $\phi$  are functions of the planar polar coordinates $r$ and $\zeta$, representing radial distance from the origin and angle to the x coordinate axis, respectively. In most systems, the ferromagnetic Hund’s coupling ($J_h>0$) aligns conduction-electron spins parallel to the local magnetization \cite{Kasuya1956,Anderson1961}.  When the spins are aligned, the coordinate space wavefunction vanishes if the two electrons come very close to one another, thereby reducing the Coulomb repulsion energy.  The strength of the skyrmion-electron interaction, $J_{h}$, typically ranges  from $1$ to $100$  meV \cite{Andrikopoulos2017,Han:2017fyd},  and plays a crucial role in transport and scattering of electrons in skyrmion-hosting environments. A minimal non-relativistic Hamiltonian describing the itinerant electrons whose spins are coupled to the localized moments is $H =\frac{\vec{ P}^2}{2m} -J_{h} \vec{M}\cdot  \vec{\sigma}$, 
where $\vec{P}$ is the momentum operator of the free electron. This can be re-expressed as  an electron effectively coupled to a uniform ferromagnet, but experiencing effective SU(2) gauge forces. We write $\psi=U^{\dagger}(\bm r)\Psi$, where  $\Psi$ is the spinor in the laboratory frame and  $\psi$ is the spinor in the locally  rotated frame,  varying from point to point in coordinate space. The SU(2) transformation matrix  $U(\bm r)$  is chosen such that  $U^{\dagger} \Big(  \bm{M}\cdot  \bm \sigma \Big)U= \sigma_{z}$, thereby aligning the local spin quantization axis with the local  magnetization $\bm M(\bm r)$. In this gauge-transformed frame, the spinor is expressed as
$\psi = \begin{pmatrix}
	\psi_{+} \\
	\psi_{-} \\
\end{pmatrix}$, where   $\psi_{+} $ and $ \psi_{-}$ correspond to electronic states with spin locally aligned and anti-aligned with  $\bm M$, respectively. This unitary matrix $U(\bm r,t)$, which  can  also vary  with time for a nonstatic magnetic background, encodes all the information about the rotation of the spin basis at each point in the spin texture.

A simple representation of the matrix field is given by $U(\bm r)=e^{i\bm \sigma\cdot \epsilon(\bm r)}$, where $\sigma$ are the Pauli matrices, and $\epsilon(\bm r)$ is a position-dependent vector encoding the spin rotation parameters.  For  rotation about a fixed axis, defined by a unit vector $\hat{n}$  and by angle $\theta$, $\epsilon(\bm{r})=\theta\hat{n}/2$. Importantly, the choice of $U(\bm r)$ is not unique. Different gauge choices correspond to different ways of defining the local spin frame while still satisfying the condition in $U^{\dagger}(\bm{M}\cdot  \bm \sigma)U= \sigma_{z}$. As we  show below, there are two topologically distinct classes of SU(2) unitary matrix fields that  can represent the same local magnetic moment field of a skyrmion. This dichotomy arises from the doubly-connected topology of the SO(3) group manifold, which describes physical  rotations in  three-dimensional space. There are two homotopically distinct classes of paths within SO(3). Paths within one class cannot be continuously  deformed to paths within the second class. The  group manifold of SO(3) can be represented as a solid ball of radius $\pi$. Each point in this manifold corresponds to  unique rotation matrix. A rotation by  an angle $\theta$ about an axis $\hat{n}$ is located at a displacement, $\theta$, from the origin of the solid ball in the direction of $\hat{n}$. Rotation by angle $\pi$  is the same as a rotation by angle $-\pi$, so the antipodal points of the solid ball are identified to be the same point. In other words, two distinct SU(2) matrices correspond to the same physical rotation. The leads to the  doubly-connected topology of the manifold. There are  two distinct classes of  matrix fields $U(\bm r)$ that can be chosen to represent the magnetic moment structure of the skyrmion. For the conventional skyrmion, as the  coordinate vector, $\vec{r}$, encircles  the skyrmion, the matrices $U_{1} (\vec{r})$  form a path, $p_1$, within SO(3) that does not cross the surface of the solid ball of radius $\pi$.  This is  depicted in Fig. \ref{so3_u1u2}. In contradistinction,  a  spin-flux carrying skyrmion is generated by the matrix field $U_2(\vec{r})$  with a path, $p_2$, within SO(3) that cannot be continuously deformed  to $p_1$ within the doubly-connected topological manifold of SO(3).  The two  paths in SO(3) are homotopically  inequivalent  because the path   $p_2$   crosses the surface of the solid ball of radius $\pi$, whereas the path   $p_1$ does not.

Applying  the SU(2) transformation field to the electron Hamiltonian, $H^{\prime}=U^{\dagger}(\bm r)H(\bm r)U(\bm r)$, modifies the momentum operator as $\vec{P} \rightarrow \vec{P} + \mathcal{\vec{ A}}$, where the emergent matrix gauge potential is given by $\mathcal{\vec{A}}=- i\hbar U^{\dagger}\bm\nabla U,$ and $\mathcal{A}_{0}=i\hbar U^{\dagger}\partial_{t} U$. In our case of static skyrmions, $\mathcal{A}_{0}=0$ since $U$ is time-independent. This leads to a synthetic matrix  magnetic field that influences electron dynamics in a manner similar to the Lorentz force in electrodynamics.  However, the  SU(2) gauge field, unlike the conventional electrodynamic vector potential, introduces a  spin-dependent force. The freedom in choosing  topologically distinct ( homotopically nonequivalent ) sets of matrices $U(\bm r)$ leads to  fundamentally and consequentially different gauge potentials. As we show below, the choice $U_2(\vec{r})$  (corresponding to  a spin-flux carrying skyrmion) leads to an SU(2) gauge field that is singular as $r \rightarrow 0$. An electron encircling this skyrmion at a large distance from its core  acquires a nontrival phase of $e^{i\pi}$  as if the skyrmion carries a magnetic solenoid containing a flux of $\pi$.  On the other hand, the choice $U_1(\bm r)$ (corresponding to a conventional skyrmion)  yields a non-singular gauge field. An electron encircling this skyrmion  acquires a trivial phase of $e^{2 i\pi}$. These two topologically distinct skyrmions exhibit identical   local magnetic moment distributions, but lead to different dynamics for the itinerant electron.

\begin{figure}
	\begin{center}
		\subfloat[\label{bzzcase1}]{
			\includegraphics[scale=0.28]{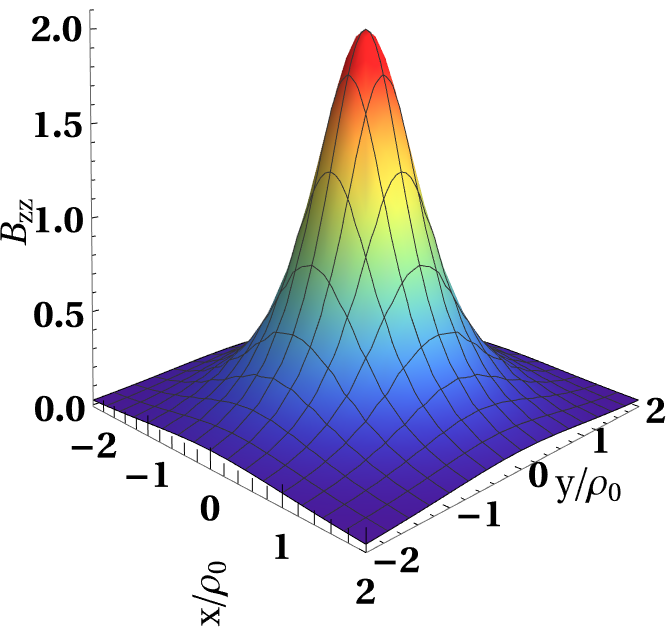}}
		\subfloat[\label{bzxcase1}]{
			\includegraphics[scale=0.28]{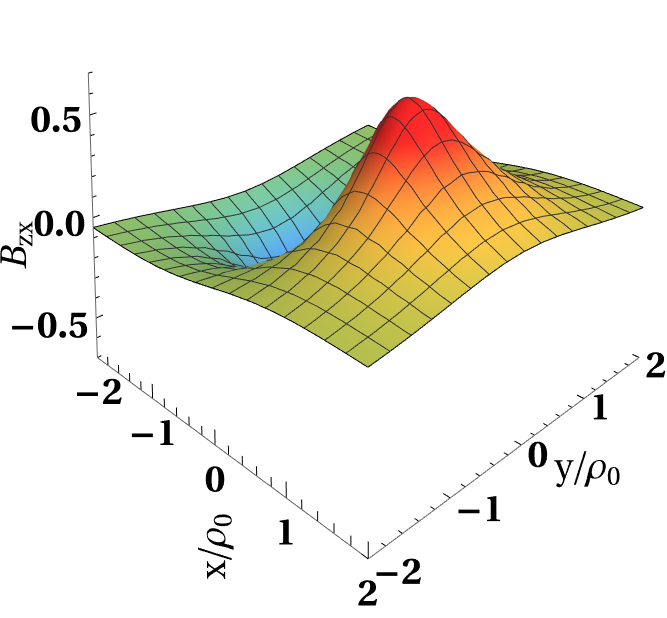}}
		\subfloat[\label{bzycase1}]{
			\includegraphics[scale=0.28]{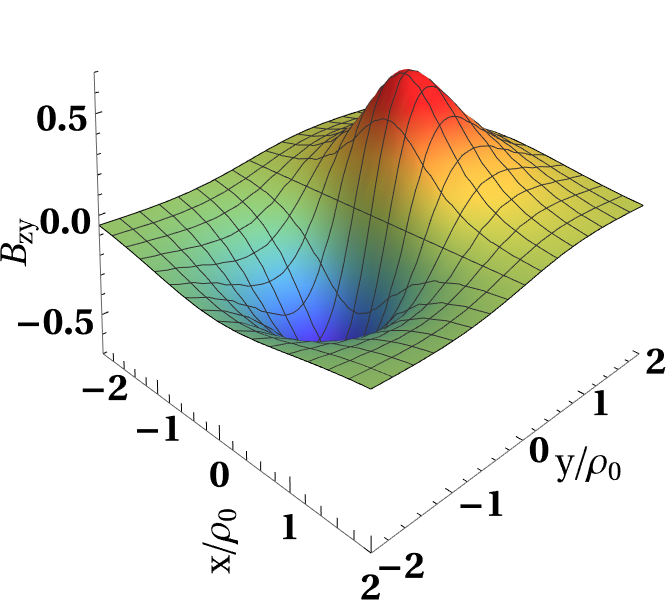}}\\ 
		\subfloat[\label{veczzcase1}]{
			\includegraphics[scale=0.28]{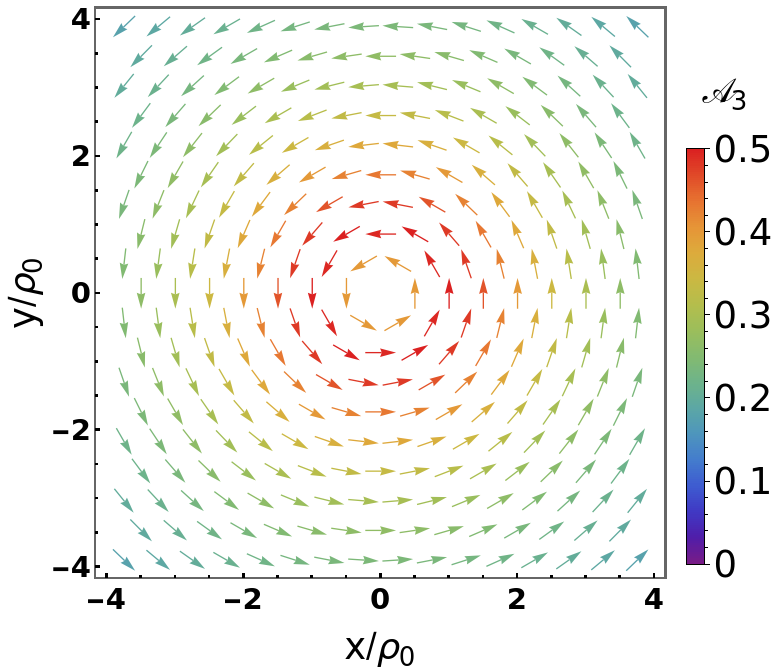}}
		\subfloat[\label{veczxcase1}]{
			\includegraphics[scale=0.28]{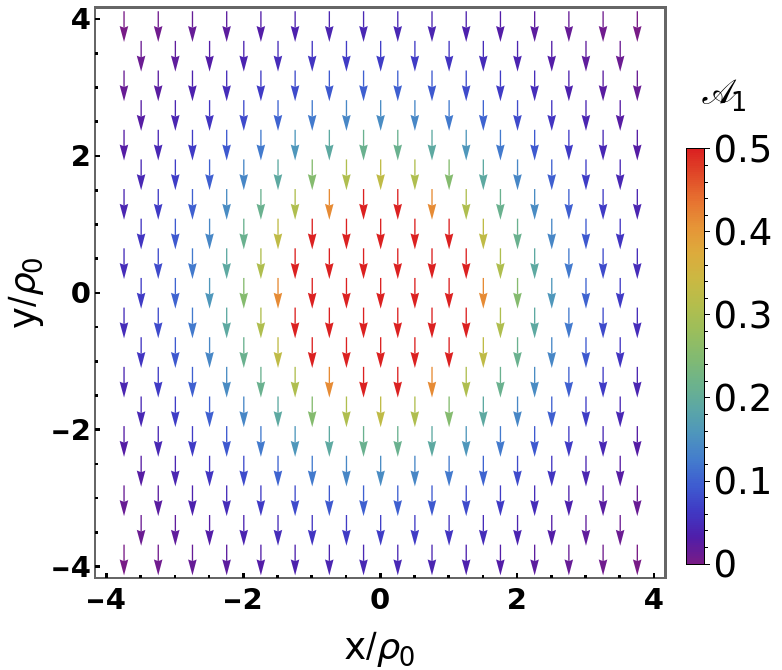}}
		\subfloat[\label{veczycase1}]{
			\includegraphics[scale=0.28]{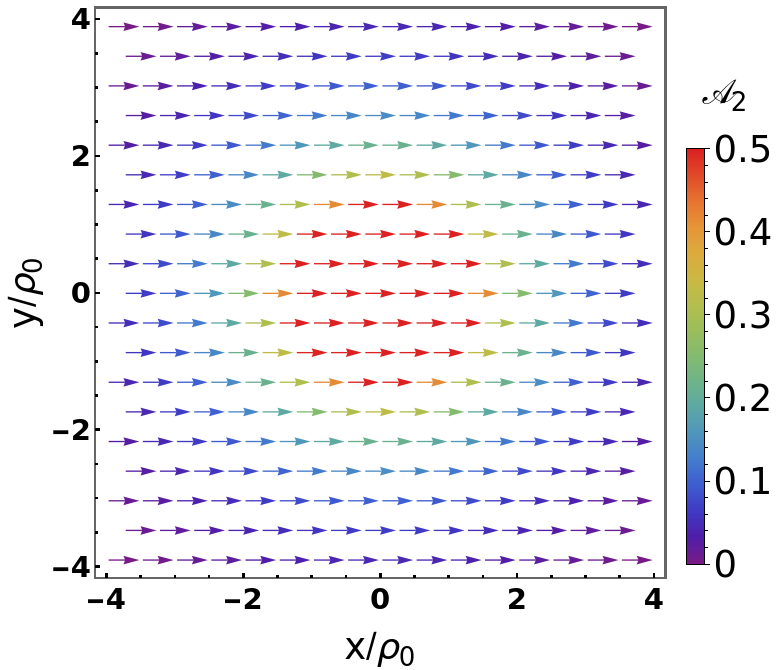}}
	\end{center}
	\caption{ Behavior of the  emergent magnetic field and vector potential for a conventional Neel skyrmion with  spin rotation $\theta(r)=  2\tan^{-1}(r/\rho_0)$, and  $U_1= \exp(-i\frac{\pi(\hat{n}\cdot\vec{\sigma})}{2})$, where $\vec{n}=(\sin\frac{\theta}{2}\cos\phi, \sin\frac{\theta}{2}\sin\phi,\cos\frac{\theta}{2})$.  Panels (a-c) display the $\sigma_z$, $\sigma_x$, and $\sigma_y$ components of the magnetic field, respectively. Panels (d-f) show the corresponding components of the vector potential. The color bars represent the magnitude of the vector potential, while the arrows indicate its direction. The emergent vector potential and magnetic field are scaled in units of $\hbar/a$ and $\hbar/a^2$, respectively, where $a$ is the lattice constant.   }
	\label{upoutwardu1}
\end{figure}

\begin{figure}
	\begin{center}
		\subfloat[\label{bzzcase2}]{
			\includegraphics[scale=0.28]{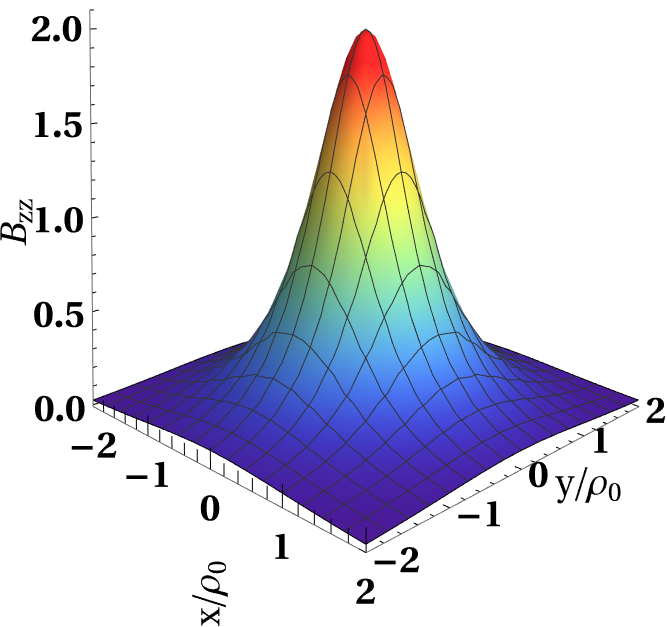}}~
		\subfloat[\label{bzxcase2}]{
			\includegraphics[scale=0.28]{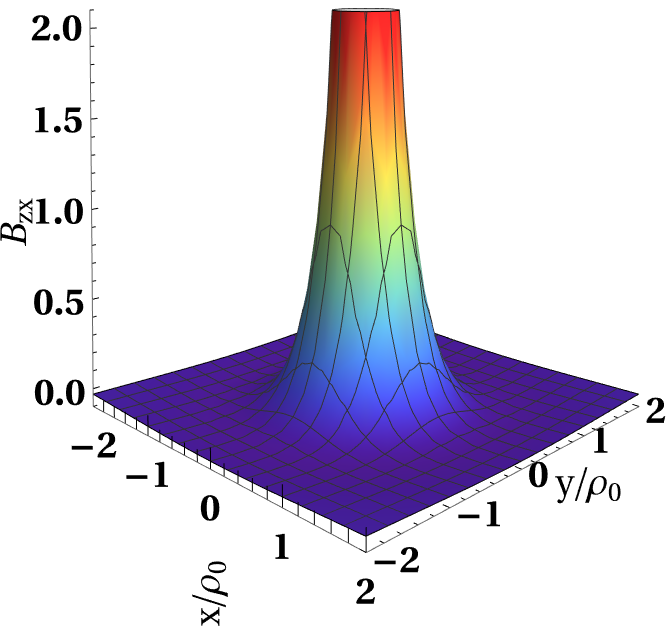}}~
		\subfloat[\label{veczzcase2}]{
			\includegraphics[scale=0.26]{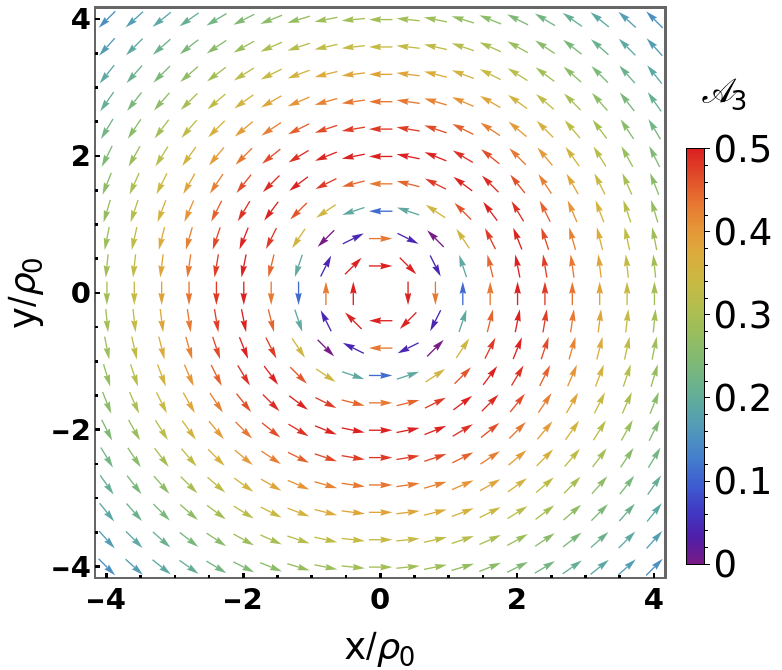}}\\
		\subfloat[\label{veczxcase2}]{
			\includegraphics[scale=0.28]{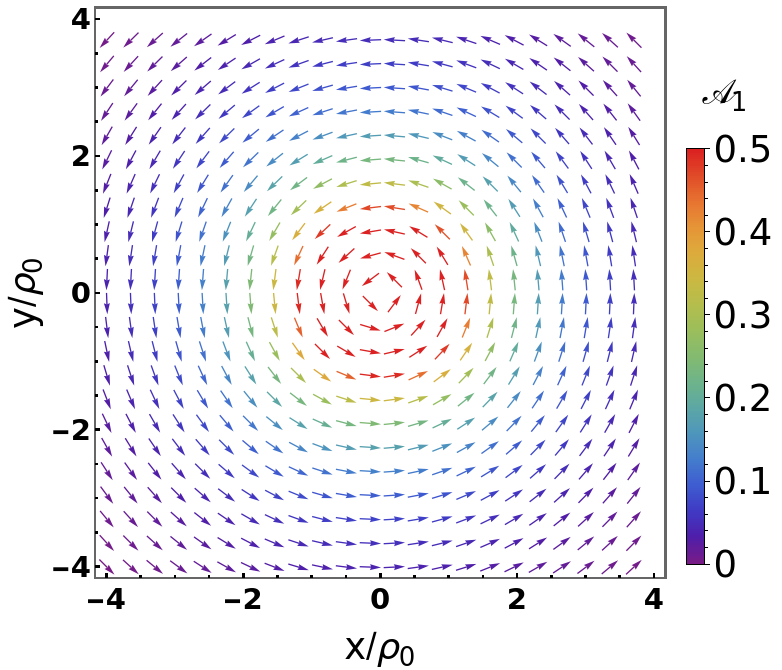}}~
		\subfloat[\label{veczycase2}]{
			\includegraphics[scale=0.28]{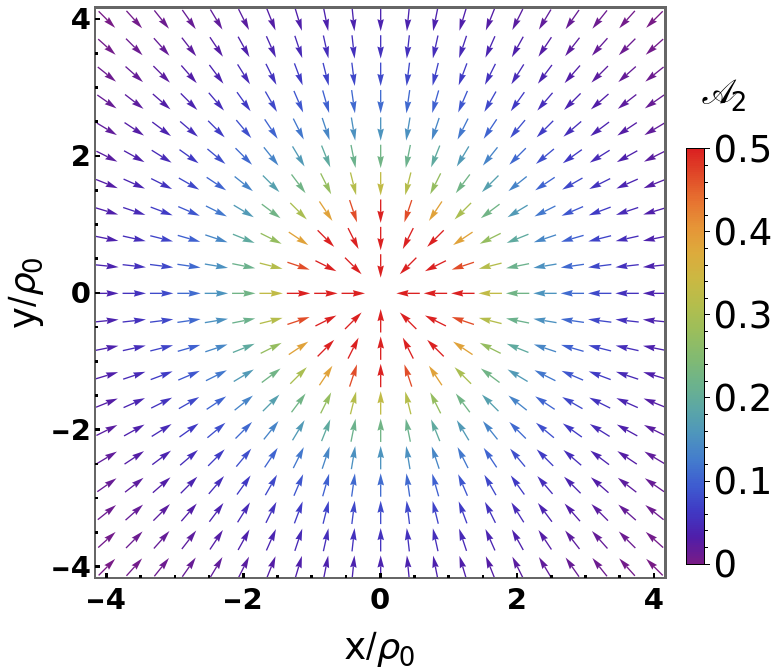}}~
		\subfloat[\label{veczz_ab_1d}]{
			\includegraphics[scale=0.25]{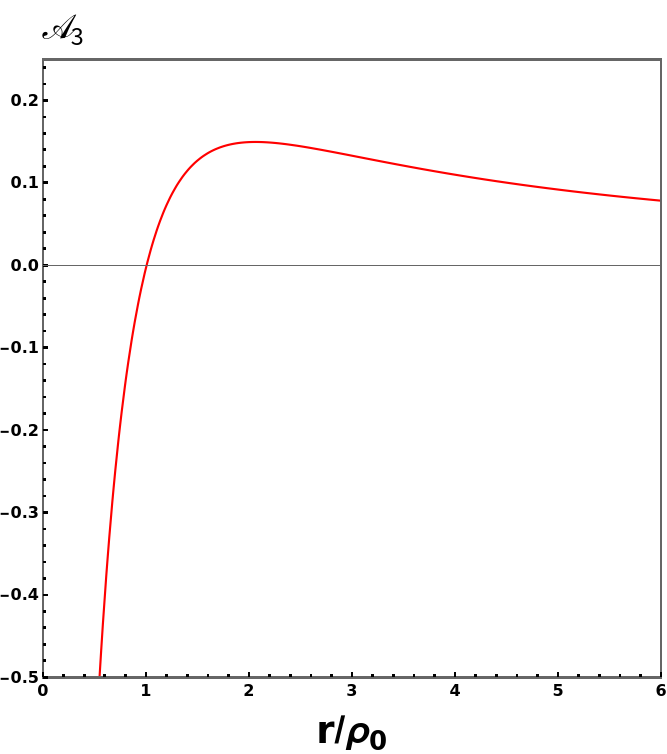}}
	\end{center}
	\caption{ The emergent  magnetic field and vector potential for a spin-flux skyrmion  with  $\theta(r)= 2\tan^{-1}(r/\rho_0)$, and  unitary matrix field $U_2 = e^{-i\frac{\sigma_{z}\zeta}{2}} e^{-i\frac{\sigma_{y}\theta(r)}{2}}$. Panels (a) and (b) show the $\sigma_z$ and $\sigma_x$ components of the  emergent magnetic field, respectively.  Notably, the $\sigma_y$ component vanishes in this configuration. Unlike $U_1$, an emergent off-diagonal magnetic field  of monopolar  form appears in the $\sigma_x$ component. Panels (c-e) display the $\sigma_z$, $\sigma_x$, and $\sigma_y$ components of the vector potential. The color bars represent the magnitude of the vector potential, while the arrows indicate its direction. Panel (f) shows the  singular radial dependence  of the $\sigma_z$ component of the vector potential, denoted as $\mathcal{A}_{3}$.   The emergent vector potential and magnetic field are scaled in units of $\hbar/a$ and $\hbar/a^2$, respectively, where $a$ is the lattice constant.}
	\label{upcenteru2}
   \end{figure}

The synthetic magnetic field of a single  $U_1(\bm r)$ skyrmion has a monopolar structure in the $\sigma_z$ component, but a dipolar structure in the $\sigma_x$  and $\sigma_y$ component  (see Fig. \ref{upoutwardu1}).  In the presence of uniform density of such skyrmions, there is an nonzero average value of the net magnetic field for the $\sigma_z$ component. However, the average magnetic field for the  $\sigma_x$  and $\sigma_y$ components vanishes due to the canceling effects of different dipoles. In contrast, the synthetic  magnetic field of a single $U_2(\vec{r})$  spin-flux skyrmion has a monopolar structure in both the $\sigma_z$  and $\sigma_x$ components (see Fig. \ref{upcenteru2}). For a uniform density of $U_2(\bm r)$ spin-flux skyrmions, there is a nonzero average value of the synthetic magnetic field for both the $\sigma_z$  and $\sigma_x$  components. This leads to  observable differences in the electron dynamics  in the presence of  $U_1(\bm r)$ vs. $U_2(\bm r)$ skyrmion densities.

In the literature, a commonly adopted form of the SU(2) unitary transformation $U_1(\bm r)$ is  \cite{Han:2017fyd,TATARA2019208}:

\begin{align}\label{case1u}
 U_1 (\bm r)& =\exp(-i\frac{\pi(\hat{n}\cdot\vec{\sigma})}{2}).
\end{align}
Here, the unit vector $\vec{n} = (\sin\frac{\theta}{2}\cos\phi, \sin\frac{\theta}{2}\sin\phi, \cos\frac{\theta}{2})$ defines the axis of rotation and the angles $\theta$ and $\phi$ are functions of 2D polar coordinates $r$ and $\zeta$. This transformation corresponds to a single rotation by angle $\pi$ about the axis $\hat{n}$, which lies halfway between the local magnetization direction and the $z$-axis. As a result, $U_1(\bm r)$ rotates the local spin texture into alignment with the $z$-axis. The rotational path traced out by such transformations within the SO(3) group manifold, as $\vec{r}$ encircles the skyrmion,  dictates key features of  the emergent gauge potential. Two example  paths, $p_1$ and $p_1^{\prime}$, within SO(3)  are depicted in Fig. \ref{so3u1}, corresponding to circles of different radii  (in coordinate space) that encircle the skyrmion.

Instead of the commonly used unitary matrix form, an alternative form of the $U(\bm r)$ matrix was introduced in  Ref. \cite{Sajeev1995}, which is  topologically distinct from  $U_1(\bm r)$  but yields the same magnetic textures. In particular, the unitary matrix $U_2(\bm r)$  follows a  sequence of two separation  rotations: 

\begin{align} \label{case2u}
U_2 (\bm r) & = \exp(-i\frac{\sigma_{z}\zeta}{2}) \exp(-i\frac{\sigma_{y}\theta ( r)}{2}).
\end{align}
The skyrmion defined by the rotation matrix field  $U_2(\bm r)$ exhibits the same topological  charge, defined by the mapping $\mathbb{R}^2  \rightarrow S^2$, as the conventional skyrmion. However $U_2(\bm r)$  defines a homotopically  distinct path within the group manifold of SO(3) from $U_1(\bm r)$ as the electron  encircles the skyrmion in coordinate space.   As we show below, this  leads to observable differences in  electron dynamics between these distinct skyrmions. The unitary matrix  product $U_2=U_2^z U_2^y$  again  aligns the electron spin with the local magnetization at each point in space. The factor, $U_2^y \equiv\exp(-i\frac{\sigma_{y}\theta}{2})$, describes the rotation by an angle  $\theta$ around the $y$-axis, aligning the spinor with the local magnetization direction in the $yz$-plane.  The factor, $U_2^z \equiv\exp(-i\frac{\sigma_{z}\zeta}{2})$,  applies a  rotation by an angle, $\phi=\zeta$, around the $z-$axis.  The path followed by  the rotations  $U_2^z$ in the group manifold of SO(3) as the skyrmion is  encircled   is depicted in Fig \ref{so3u2}. The skyrmion involving $U_2^z$  exhibits a  consequential ``spin-flux" of  $\pi$ causing spin-dependent   destructive wave interference of  two  electron trajectories  passing either side of the skyrmion core.

The  effective gauge potential $\mathcal{\vec{A}}$ modifies the electron motion similarly to a real magnetic vector potential except in a spin-dependent manner. This can be described by  the emergent field $\vec{B} =\bm \nabla \times \mathcal{\vec{A}}$. One  consequence of this emergent field is the topological Hall effect (THE), where electrons experience a transverse  Lorentz force  due to their interaction with the skyrmion. The transformed Hamiltonian  $H^{\prime}=  \frac{1}{2m} ( \bm P +\mathcal{ \bm A})^2  -J_{h}  \sigma_z$ now includes an emergent  SU(2) gauge potential $\mathcal{ \bm A}$. Depending on the homotopically distinct choices of the unitary matrix fields, $U_1$  or $U_2$, the structure of the emergent gauge potential and its associated physical consequences differ. In the next section, we  analyze the explicit forms of the SU(2) gauge potentials produced by a single skyrmion. In particular, the skyrmion created by $U_{2}(\bm r)$ exhibits monopolar  off-diagonal  components in its emergent magnetic field and a vector potential similar to that of a magnetic solenoid carrying a magnetic flux of $\pi$.  In our analysis, we treat the skyrmion as a static texture.

\subsection{Synthetic Gauge Field Due to Conventional Skyrmions}\label{case1}

The  local  magnetization  $\bm M = (\sin\theta\cos\phi, \sin\theta\sin\phi,\cos\theta) $, is defined at each point  $\bm r = (x,y)\equiv (r\cos\zeta, r\sin\zeta)$, where $r$ is the radial distance from the origin and $\zeta$ is the angle with respect to the x-axis.  The angles, $\theta$ and $\phi$ are functions of  $r$ and $\zeta$. For the $U_1$ unitary matrix, the emergent gauge vector potential  is given by  (see Appendix \ref{emergentu1})

\begin{align}
 \mathcal{\vec{A}} =\frac{\hbar}{2}\Big[ \theta_r(\sigma_{y}\cos\phi -\sigma_{x}\sin\phi)\hat{r}  +   \hat{\zeta} \frac{2 \phi_\zeta\sin \frac{\theta}{2} }{r} \Big(\sigma_{z} \sin \frac{\theta}{2} -\cos \frac{\theta}{2} (\sigma_{x}\cos\phi +\sigma_{y}\sin\phi)\Big)\Big],
\end{align} 
where $\theta_r \equiv \partial \theta/\partial r$ and $\phi_\zeta \equiv \partial \phi/\partial \zeta$. The first term of the gauge potential is oriented along  the unit vector $\hat{r}$ in  the radial direction, while the second term is aligned the unit vector $\hat{\zeta}$ in the angular direction.  We consider a skyrmion whose central spin is aligned along the $+z$ direction, with a  radially outward rotation described  by $\theta(r)=2\tan^{-1}(r/\rho_0)$.  In all  our calculations, we consider only Néel-type skyrmions,  for which $\phi = \zeta$ \cite{Fert2013,sandip2019}. In this case, the detailed form of the emergent vector potential and the corresponding emergent magnetic field are (see Appendix \ref{appendixu1}  for  details):

\begin{align}  
  \label{veccase1}& \mathcal{\vec{A}} =  \frac{\hbar}{2}\Big[ \frac{2\rho_0}{r^2 +\rho_0^2}(\sigma_{y}\cos\zeta -\sigma_{x}\sin\zeta)\hat{r}  +   \hat{\zeta} \Big(\sigma_{z}  \frac{2r}{r^2 +\rho_0^2} - \frac{2\rho_0 }{r^2 +\rho_0^2}(\sigma_{x}\cos\zeta +\sigma_{y}\sin\zeta)\Big)\Big] \equiv \sigma_x \mathcal{\vec{A}}_{1}  +\sigma_y \mathcal{\vec{A}}_{2} +\sigma_3 \mathcal{\vec{A}}_{3}\\ &   \label{magcase1}
B_z = \hbar \Bigg[ \sigma_z \frac{2\rho_0^2}{(r^2 +\rho_0^2)^2}  +\frac{2 r\rho_0}{(r^2 +\rho_0^2)^2} (\sigma_x \cos\zeta +\sigma_y \sin\zeta)\Bigg]  \equiv \sigma_z B_{zz}  +\sigma_x B_{zx} +\sigma_y B_{zy}\,.
\end{align}
The behavior of  $\sigma_z$, $\sigma_x$ and $\sigma_y$ components of the vector potential are shown in Figures  \ref{veczzcase1},  \ref{veczxcase1} and  \ref{veczycase1}, respectively.  Here, the arrows indicate the direction of the vector potential at each point, while the colored bars represent its magnitude. As shown in Figure \ref{veczzcase1}, the  $\sigma_z$ component of the vector potential is maximal at the skyrmion center and decays as  $1/r$ at large distances. In contrast, $\sigma_x$ and $\sigma_y$ components, shown in figures \ref{veczxcase1} and  \ref{veczycase1},  decay  more rapidly, following a   $1/r^2$  dependence.  Additionally, the maximum value of the  $\sigma_z$ component  is approximately  $1.4$  that of  the  other two components.  Since the  $\sigma_z$ component of $\mathcal{A}$ is aligned along the azimuthal direction, it contributes directly to the magnetic flux.  On the other hand,  $\sigma_x$ and $\sigma_y$ components  are aligned  purely along the $y-$ and $x-$directions, respectively and do not  produce  net magnetic flux. The total magnetic flux enclosed  within a circle of radius $\mathcal{R}$ is given in Eq. \ref{afluxcase1}. In the limit $  \mathcal{R} \rightarrow \infty$,  this total flux approaches $2\pi$. The flux experienced  by an up-spin electron is opposite to that of a down-spin electron.   Since the vector potential  lies in the x-y plane  it  generates only a $z-$component of  synthetic SU(2) magnetic field.

The explicit form of this fictitious magnetic field is given in equation \ref{magcase1}. The $\sigma_z$ component of the magnetic field, denoted $B_{zz}$, is maximum at the center of the skyrmion and decays as $1/r^4$ with distance. The behavior of $B_{zz}$ in the $x-y$ plane is shown in Figure \ref{bzzcase1}. Among all components, $B_{zz}$ has the highest magnitude-approximately four times larger than the $B_{zx}$ and $B_{zy}$ components depicted  in  Figures \ref{bzxcase1} and \ref{bzycase1}, respectively. The $B_{zx}$ component exhibits an antisymmetric dipolar structure along the $x$-axis resulting in zero net magnetic flux over the skyrmion area. The $B_{zy}$ component shows a similar  dipolar profile along the $y$-axis, also contributing no net  magnetic flux.  For this conventional (nonsingular) skyrmion:

\begin{align} \label{afluxcase1}
&\oint_{C} \mathcal{\vec{A}}\cdot d\vec{l}\equiv \int_{-\pi}^{\pi} \mathcal{\vec{A}} (\mathcal{R},\zeta)\cdot \hat{\zeta}\mathcal{R}d\zeta =2\pi\hbar\Big(1-\frac{\rho_0^2}{\mathcal{R}^2 +\rho_0^2} \Big)  \sigma_{z}=2\pi\hbar \sigma_{z}\big|_{\mathcal{R} \rightarrow \infty} \\ 
& \int_{S} B_z dxdy \equiv \int_{0}^{\mathcal{R}}  \int_{-\pi}^{\pi} B_z (r,\zeta) rdrd\zeta =2\pi\hbar\Big(1-\frac{\rho_0^2}{\mathcal{R}^2 +\rho_0^2} \Big)\sigma_{z}  =2\pi \hbar \sigma_{z} \big|_{\mathcal{R} \rightarrow \infty}
\end{align}
Here, the line integral is taken along a circular contour $C$ of radius $\mathcal{R}$,  and the surface integral extends over the circular disk $S: 0\leq r\leq \mathcal{R}, -\pi\leq \zeta \leq \pi$

\subsection{Singular Gauge Field Due to Spin-Flux Skyrmions}
The general expression for the SU(2) gauge potential associated with a $U_2$ spin-flux skyrmion is given by  (see Appendix \ref{emergentu1})

\begin{align}
\mathcal{\vec{A}} =\frac{\hbar}{2}\Big[- \theta_r\sigma_{y}\hat{r}  +   \frac{\hat{\zeta} \phi_\zeta}{r} \Big(-\sigma_{z} \cos\theta + \sigma_{x}\sin \theta\Big)\Big].
\end{align}
We again consider a representative up-center skyrmion configuration defined by $\theta(r) = 2\tan^{-1}(r/\rho_0)$,  and $\phi=\zeta$. For this choice, the explicit analytical forms of the emergent gauge potential and the corresponding  SU(2) magnetic field  exhibit singularities as $r \rightarrow 0$ (see Appendix \ref{appendixu2}  for  details):

\begin{align}  \label{veccase2}
& \mathcal{\vec{A}} = \frac{\hbar}{2}\Big[- \frac{2\rho_0}{\rho_0^2 +r^2}\sigma_{y}\hat{r}  +   \hat{\zeta} \Big(-\sigma_{z} \frac{\rho_0^2 -r^2}{r(\rho_0^2 +r^2)} + \sigma_{x}\frac{2\rho_0}{\rho_0^2 +r^2}\Big)\Big]\\ & \label{magcase2}
B_z  =  \hbar \Bigg[ \sigma_z \frac{2\rho_0^2}{(\rho_0^2 +r^2)^2}  +\frac{\rho_0(\rho_0^2-r^2)}{r(\rho_0^2 +r^2)^2}\sigma_x \Bigg]
\end{align} 
The behaviors of the  $\sigma_z$, $\sigma_x$ and $\sigma_y$ components of the vector potential are illustrated  in Figures  \ref{veczzcase2},  \ref{veczxcase2} and  \ref{veczycase2}, respectively. Here, the arrows indicate the direction of the vector at each point,  and  the colored bars represent the magnitude of the   vector potential. The $\sigma_z$ component of the vector potential  exhibits a $1/r$ singularity at the center of the skyrmion and  decays as $1/r$ for larger. The other two components are nonsingular as $r \rightarrow 0$ and  decay as $1/r^2$ at large distances.  Both $\sigma_z$  and  $\sigma_x$ components are aligned along the azimuthal direction, contributing to the total magnetic flux. The explicit form of the flux inside a circle of radius $\mathcal{R}$ is given in Eq. \ref{afluxcase2}. Unlike the case discussed in Section \ref{case1}, where the magnetic flux originates solely from the $\sigma_z$ component, the $U_2$  skyrmion exhibits  flux contributions from both $\sigma_z$ and $\sigma_x$. As shown in Fig. \ref{veczz_ab_1d}, the $\sigma_z$ component of the azimuthal vector potential (see Eq. \ref{veccase2}) changes sign at sign at $r = \rho_0$. For $r<\rho_0$,  this component is negative, resulting in a clockwise circulation of the vector field. For $r>\rho_0$, it becomes positive, reversing the rotation direction to counterclockwise (see  Fig. \ref{veczzcase2}). This is a distinctive behavior for  $U_2$-skyrmion not present in the  $U_1$-skyrmion. In the limit $\mathcal{R} \rightarrow \infty$, the total magnetic  flux approaches $2\pi$. However, the line integral of the vector potential approaches  $\pi$ leading to distinct phase changes from the $U_1$-skyrmion as the electron encounters the  $U_2$-skyrmion. An electron encircling the $U_2$ skyrmion  acquires a non-trivial phase factor of $e^{i\pi}$. For this reason, we refer to the $U_2$ skyrmion as a ``spin-flux" skyrmion. The usual Stoke's theorem, relating magnetic flux to the line integral of the vector potential, is inapplicable due to the  singular form of the gauge field as $r \rightarrow 0$.

The explicit form of the fictitious magnetic field for the $U_2$ spin-flux skyrmion is given in Eq. \ref{magcase2}. The $\sigma_z$ component of the magnetic field, denoted by $B_{zz}$, reaches its maximum at the center of the skyrmion, as shown in Figure \ref{bzzcase2}. This behavior closely resembles that observed for the $U_1$ skyrmion discussed earlier.  In contrast, the $B_{zx}$ component of the $U_2$ spin-flux skyrmion, shown in Figure \ref{bzxcase2}, exhibits a unique  structure. Unlike the dipolar pattern observed in the $U_1$-skyrmion, the $B_{zx}$ field now displays a  singular monopole-like structure centered at the skyrmion core and also decays as $1/r^3$  for large  distances. For the (singular) spin-flux skyrmion:

\begin{align} 
& \int B_z dxdy  =2\pi\hbar\Big[\Big(1-\frac{\rho_0^2}{\mathcal{R}^2 +\rho_0^2} \Big)  \sigma_{z} +\frac{\mathcal{R}\rho_0}{\mathcal{R}^2 +\rho_0^2}   \sigma_{x}\Big]\big|_{\mathcal{R} \rightarrow \infty} =2\pi\hbar \sigma_{z} \\ & 
\oint \mathcal{\vec{A}}\cdot d\vec{l}  = 2\pi \hbar\Big[\Big(\frac{1}{2}-   \frac{\rho_0^2}{\mathcal{R}^2 +\rho_0^2} \Big)\sigma_{z} + \frac{\mathcal{R}\rho_0}{\mathcal{R}^2 +\rho_0^2}\sigma_{x}\Big]\big|_{\mathcal{R} \rightarrow \infty}  \label{afluxcase2} = \pi \hbar \sigma_{z} 
\end{align}

\begin{figure}
	\begin{center}
		\subfloat[\label{bzzcircle3d}]{
			\includegraphics[scale=0.26]{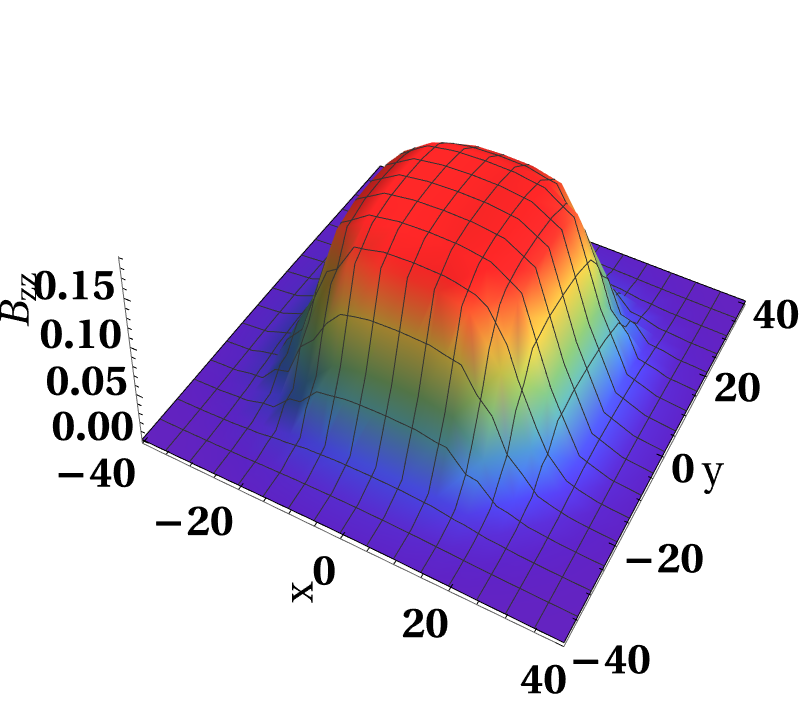}}~
		\subfloat[\label{bzxu1circle3d}]{
			\includegraphics[scale=0.26]{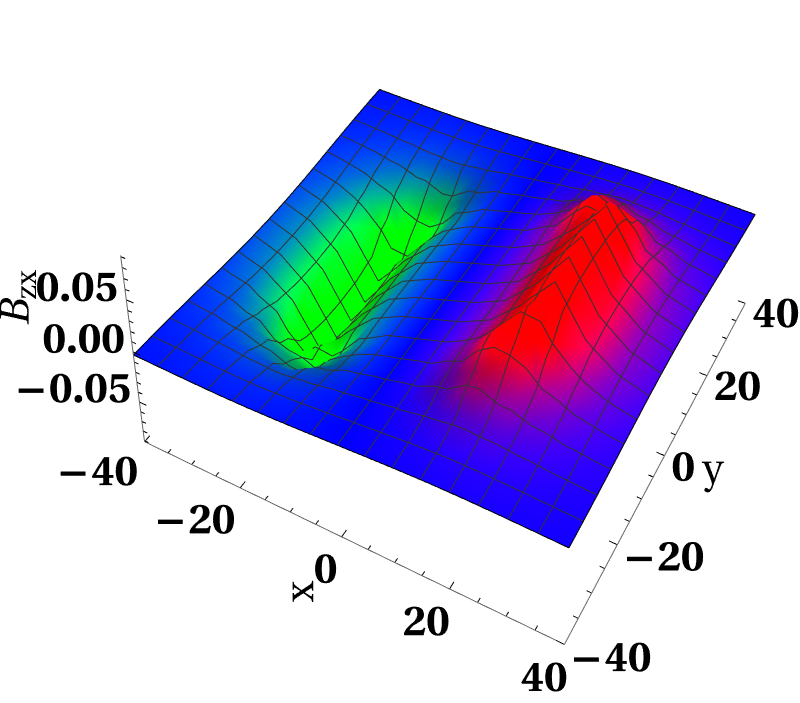}}~
		\subfloat[\label{bzxu2circle3d}]{
			\includegraphics[scale=0.26]{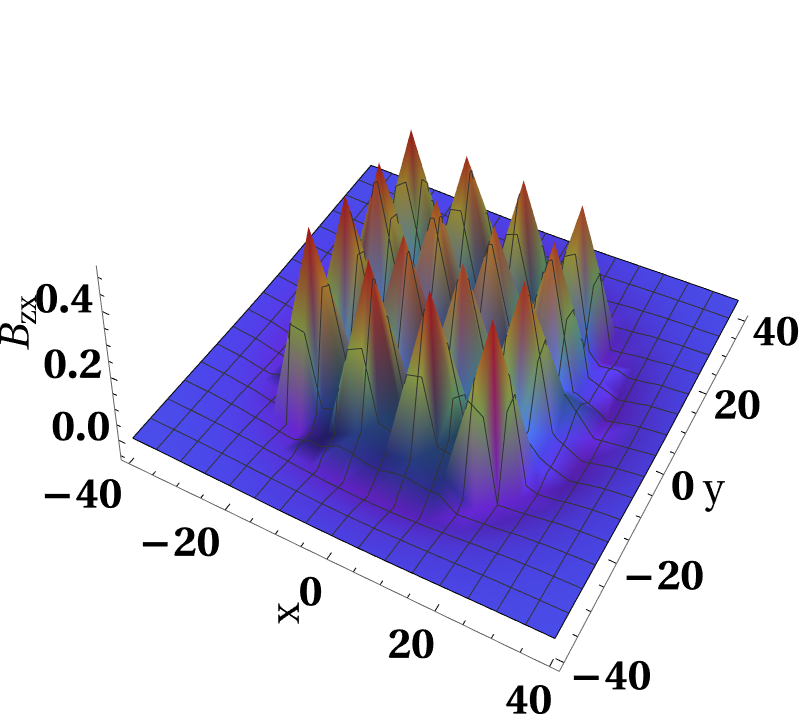}}
	\end{center}
	\caption{Spatial profiles of components of the  emergent magnetic field, for a lattice of skyrmions with core radius $\tilde{\rho_0}=5a$,  within the large circular region of radius $\tilde{r}=20a$. Panel (a) shows the distribution of the $B_{zz}$ component in the $x-y$ plane  for a lattice of either $U_1$ or $U_2$ skyrmions. For both skyrmion lattices, $B_{zz}$ remains nearly uniform throughout the interior but drops sharply near the boundary. Panel (b) presents the spatial variation of $B_{zx}$ for the $U_1$ skyrmion lattice. As expected, $B_{zx}$ is nearly zero across the central area but becomes nonzero near the boundary and displays a dipolar-like structure. In contrast, panel (c) illustrates the behavior of $B_{zx}$ for the $U_2$ skyrmion lattice. This has monopolar  peaks at the center of each skyrmion,  leading to a significant average value over the large circular area.}
	\label{appexdix1}
\end{figure}

\subsection{ Average field of a  skyrmion lattice}

\begin{figure}
	\begin{center}
		\subfloat[\label{bzzden}]{
			\includegraphics[scale=0.24]{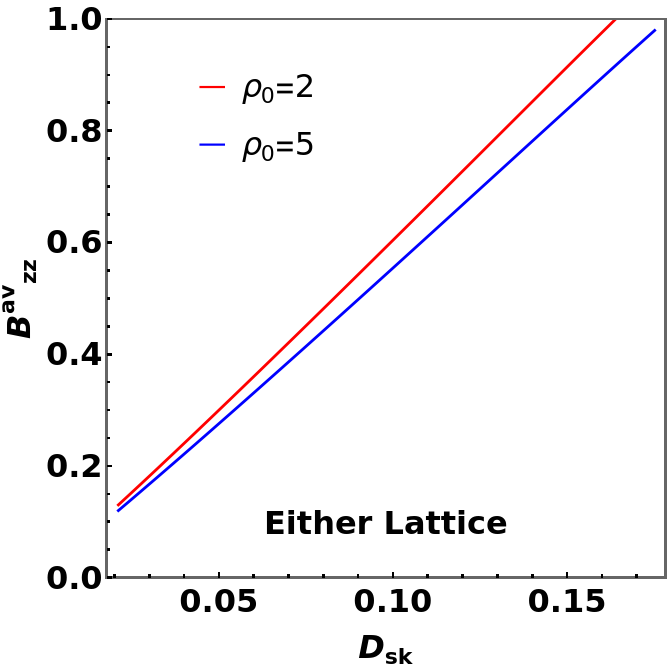}}
		\subfloat[\label{bzzrho}]{
			\includegraphics[scale=0.255]{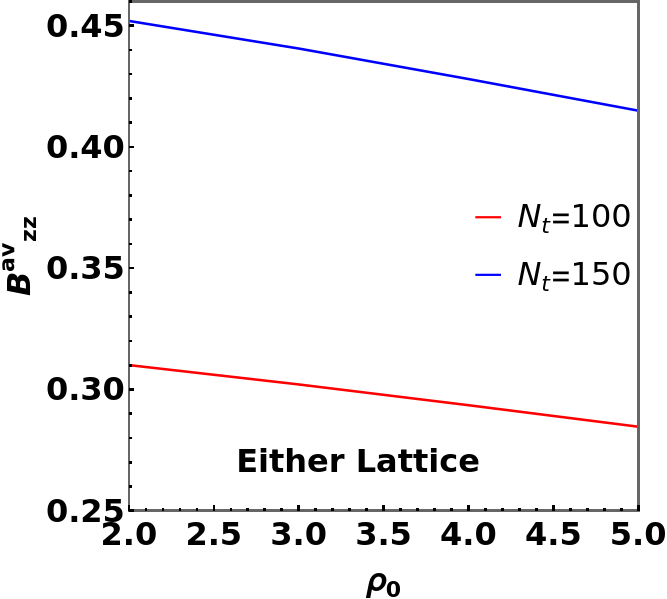}}
		\subfloat[\label{bzxden}]{
			\includegraphics[scale=0.24]{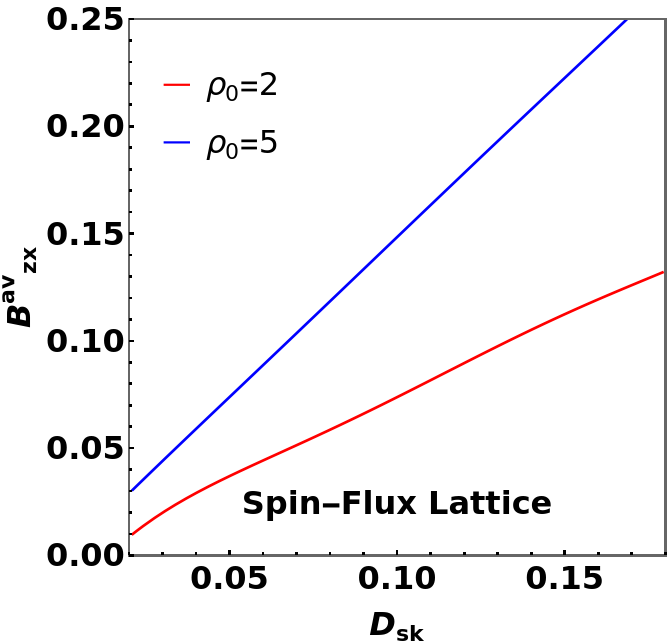}}
		\subfloat[\label{bzxrho}]{
			\includegraphics[scale=0.24]{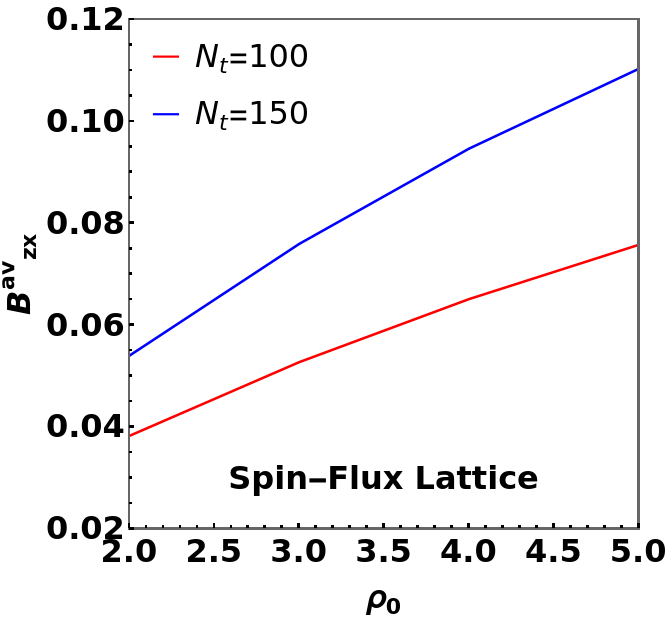}}
	\end{center}
	\caption{Panels (a) and (b) show the behavior of the averaged magnetic field component $B_{zz}^{\text{av}}$ as a function of skyrmion density, $D_{\text{sk}}$, and skyrmion radius $\rho_0$, respectively.   $B_{zz}^{\text{av}}$ remains the same  for both $U_1-$skyrmion and spin-flux $U_2-$skyrmion lattices.	Panels (c) and (d) display the behavior of the averaged   off-diagonal field  component $B_{zx}^{\text{av}}$ for $U_2-$skyrmion lattice, as a function of $D_{\text{sk}}$ and $\rho_0$, respectively.  The amplitude of $B_{zz}^{\text{av}}$ is  larger than that of $B_{zx}^{\text{av}}$, due to their distinct functional forms, as described in Eqs. \eqref{magcase1} and  \eqref{magcase2}. The dependence of $B_{zz}^{\text{av}}$ and $B_{zx}^{\text{av}}$ on  $\rho_0$ is qualitatively different  as seen by comparing  panels (b) and (d). For  the $U_1-$skyrmion lattice, the average values of the off-diagonal field components are nearly zero. In all cases,  the region containing skyrmions   has a radius of $25a$, where $a$  is the  lattice spacing between magnetic moments. }
	\label{bzavr}
\end{figure}

\begin{figure}
	\begin{center}
		\subfloat[\label{bzzall}]{
			\includegraphics[scale=0.29]{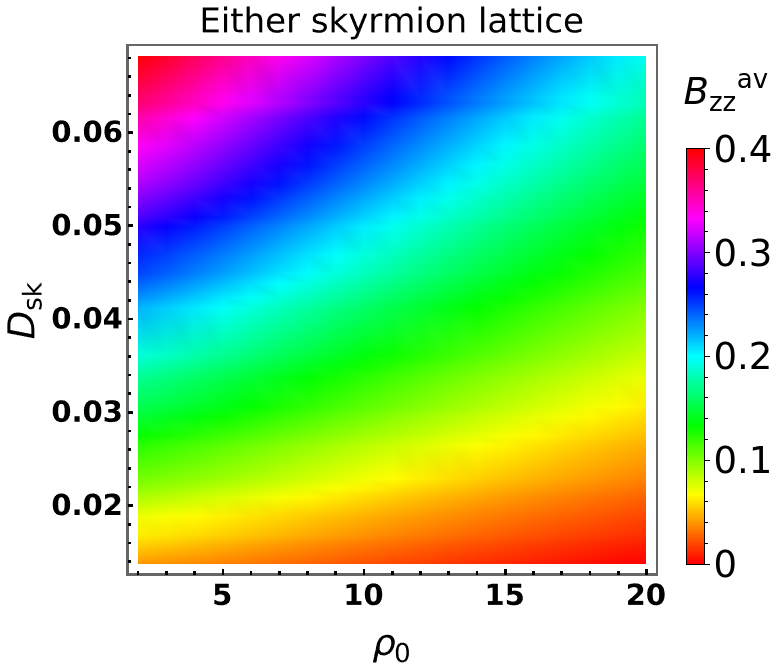}}~
		\subfloat[\label{bzxall}]{
			\includegraphics[scale=0.29]{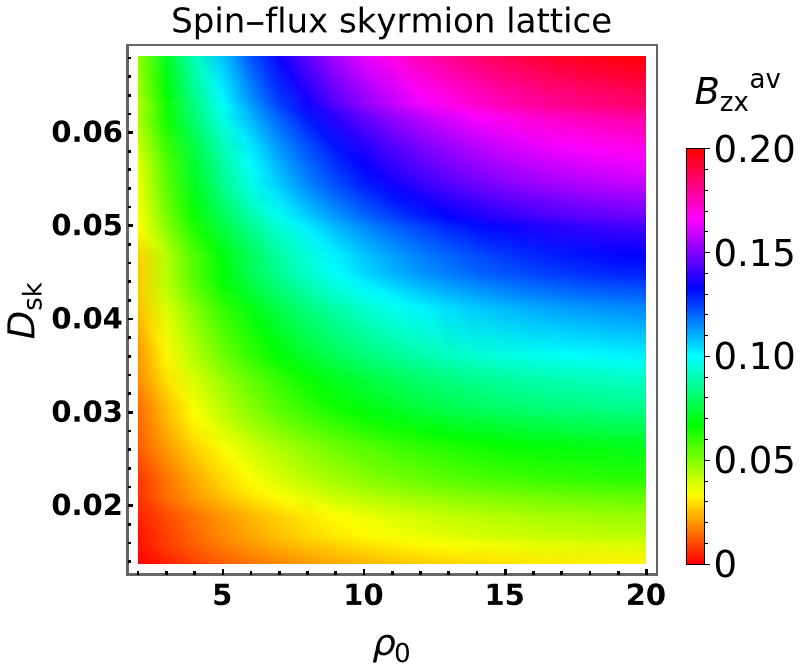}}~
		\subfloat[\label{circleskyrmion}]{
			\includegraphics[trim={0 -3cm 0 0cm}, scale=0.28]{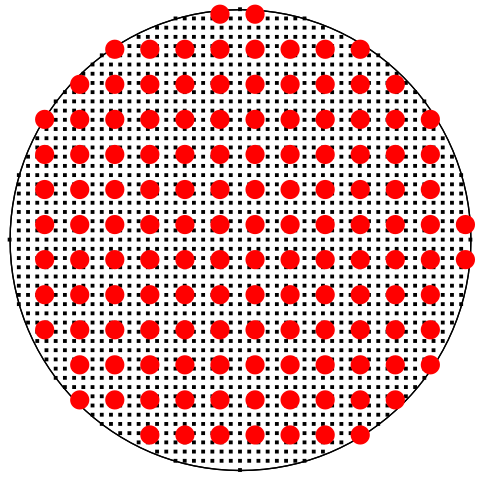}}
	\end{center}
	\caption{Panel (a) depicts the averaged magnetic field component $B_{zz}^{\text{av}}$ ( identical for both  $U_1-$skyrmions and  $U_2$ spin-flux skyrmions) as a function of  skyrmion density $D_{\text{sk}}$ and skyrmion radius $\rho_0$. The color bar indicates the  magnitude  of $B_{zz}^{\text{av}}$.  Panel (b) displays the distribution of the $B_{zx}^{\text{av}}$ component   for the $U_2$ spin-flux skyrmion in the same parameter space. Clearly, $B_{zz}^{\text{av}}$ increases with skyrmion density but decreases with increasing skyrmion radius.  In contrast, $B_{zx}^{\text{av}}$ is non zero and large only in $U_2$-skyrmions. This shows a monotonic increase with both skyrmion density and radius.  Panel (c) shows a schematic  representation of skyrmions arranged  in a square lattice within a large circle of radius $\tilde{r}_0 = 25a$. The red dots denote individual skyrmions, while black dots mark the underlying lattice sites of individual magnetic moments. The radius of each skyrmion is set to $\tilde{\rho}_0 = 2a$.}
\end{figure}

We  now consider a periodic arrangement of skyrmions confined within a large circular region of radius $r_0$, as illustrated in Figure \ref{circleskyrmion}.  Here, we assume that the presence of multiple skyrmions can be represented by a linear superposition of their emergent gauge fields. The total emergent magnetic field is obtained by summing the contributions from all individual skyrmions within this region. For the $\sigma_z$ component, the total field  for both the $U_1-$skyrmion and the $U_2-$skyrmion is given by:

\begin{align}
B_{zz}^{total}  =\sum_{i}B_{zz}^{i}&= \sum_{i}\frac{2\rho_0^2 \hbar}{\Big[(x-x_0^i)^2 +(y-y_0^i)^2+\rho_0^2\Big]^2} 
\end{align}
Here, $(x_0^i, y_0^i)$ are the coordinates of the center of the $i^\text{th}$ skyrmion, and $\pi r_0^2$ is the total area of the skyrmion region. The total number of skyrmions is $N_t$, and the skyrmion density is defined as $D_{sk} = \frac{N_t}{\pi r_0^2}$.  To obtain the average magnetic field, we integrate the total field over the system area. For the $\sigma_z$ component, the average is given by $	B_{zz}^{av} =\frac{1}{\pi r_0^2} \int B_{zz}^{total}dx dy$. The average values of the other components, $B_{zx}^{\text{av}}$ and $B_{zy}^{\text{av}}$, can be computed similarly and we denote the matrix average as  $ B_z^{av} =B_{zx}^{av}\sigma_x +B_{zy}^{av}\sigma_y+B_{zz}^{av}\sigma_z$. In  what follows, we generalize $B_{z}^{av}$ to represent a configuration average,  $[B_z]_c$, over different skyrmion coordinate  locations  (not restricted to those depicted in Fig. \ref{circleskyrmion}). We assume that  $[B_z]_c=B_z^{av}$. The average emergent magnetic field components, $B_{zz}^{av}$ and $B_{zx}^{av}$, are functions of skyrmion density and radius. As shown in Figure \ref{bzzden}, $B_{zz}^{av}$ increases with skyrmion density for both $U_1$- and $U_2$-type skyrmions. Conversely, Figure \ref{bzzrho} shows that $B_{zz}^{av}$ decreases with increasing skyrmion core radius $\rho_0$. This   highlights the competing roles of skyrmion size and density in determining the  spatially averaged field.

The off-diagonal components  $B_{zx}$  and $B_{zy}$ of the emergent magnetic field of a conventional skyrmion ($U_1$) exhibit  dipole-like structures oriented along the x- and y-directions, respectively, as illustrated in Figure \ref{upoutwardu1}. Consider a crystal of  skyrmions filling a large circular region (see figure \ref{circleskyrmion}). In such a configuration, the dipole field of one skyrmion can overlap with that of a nearby skyrmion, leading to partial or complete cancellation  of the local field in the interior region.  In any event, the spatial averages the off-diagonal components of the emergent field of the $U_1-$skyrmion crystal cancel out in the bulk of the system. However, this cancellation does not hold at the boundaries  if the dipolar fields are aligned in one direction.

This situation is quite different for a lattice of $U_2$-skyrmions. Figure \ref{bzxden} show  that the average emergent field component $B_{zx}^{av}$ increases with skyrmion density. As shown in Figure~\ref{bzxrho}, $B_{zx}^{av}$ initially increases with skyrmion radius  \(\rho_0\), but eventually saturates. This behavior can be understood from the  monopolar nature of the emergent field of a single skyrmion $B_{zx}(r) \propto \frac{\rho_0 (\rho_0^2 - r^2)}{r (\rho_0^2 + r^2)^2},$ which is singular at  the skyrmion center and decays as \(1/r^3\) for \(r \gg \rho_0\).  This  distinctive behavior  provides a framework  for interpreting  certain experimental results on the Hall conductivity  as function of applied magnetic field, as we discuss below  (see Fig. \ref{appexdix3}).

In summary, for the $U_1$-skyrmion, the spatially-averaged emergent magnetic field component $B_{zz}^{av}$  increases with skyrmion density but decreases with skyrmion core radius, while the off-diagonal components $B_{zx}^{\text{av}}$ and $B_{zy}^{\text{av}}$ are negligibly small. In contrast, for the $U_2$ spin-flux skyrmion, both $B_{zz}^{\text{av}}$ and $B_{zx}^{\text{av}}$  are significant,  with $B_{zx}^{\text{av}}$   increasing  with skyrmion core radius.   Thus, the averaged emergent matrix magnetic fields for a finite density of skyrmions takes the form:

\begin{align}
	B_z^{av} =
	\begin{cases} 
		B_{zz}^{av}\sigma_z , & \text{for $U_1$ (Conventional skyrmions) }  \\
		B_{zz}^{av}\sigma_z +B_{zx}^{av}\sigma_x , & \text{for $U_2$ (Spin-flux skyrmions)}
	\end{cases}
\end{align}
This contributes  to different dynamics for an itinerant electron encountering a region of $U_1-$skyrmions  vs. $U_2-$skyrmions   even in a semi-classical picture, as demonstrated in Section \ref{spindrude}.

\section{Spin-dependent Drude theory}\label{spindrude}

We  now consider the semi-classical dynamics of an electron subject to the emergent matrix-valued magnetic field generated by a finite density of skyrmions. In this framework, fluctuations of the emergent field around its average value are treated as an additional scattering source, contributing to an effective relaxation time $\tau$. The resulting transport behavior is described using a spin-dependent generalization of the Drude model \cite{AshcroftMermin1976}. We start from an effective Heisenberg equation of motion for the electron momentum operator:

\begin{align}\label{drudefirst}
	m \dot{\vec{v}} & = -q\vec{E}\sigma_{0} - m\frac{\vec{v}}{\tau} - B_{z}(\vec{v}\times \hat{z}).
\end{align}
Here $\vec{v}=\hat{x}\dot{x}+\hat{y}\dot{y}$ is the velocity operator, which includes the identity matrix, $\sigma_0$, in spin space.  Here, $\vec{E}$ is the  physical external electric field, and $B_{z}$ represents the spin-dependent emergent magnetic field arising from a finite density of stationary skyrmions. The term $B_z(\vec{v}\times \hat{z})$  is a spin-dependent Lorentz-like force due to this emergent field, while the damping term $m\frac{\vec{v}}{\tau}$ phenomenologically accounts for momentum relaxation from impurities, emergent-field fluctuations and other sources of random scattering.  A detailed derivation is presented in Appendix~\ref{homogeneousderivation}.  The first term on the right-hand side corresponds to the force exerted by the external electric field, acting identically on all spin states. The Drude-like equation \eqref{drudefirst} gives rise to  spin-selective transport determined by the specific form of the conduction electron spin density matrix.

We now take the ensemble  average and quantum expectation value of Eq.~\ref{drudefirst} under steady-state conditions, where the average acceleration vanishes, i.e., $\langle \dot{\vec{v}} \rangle_{\rm ens} = 0$. Assuming a uniform applied electric field along the $x$-axis, the spin-dependent Drude equation reduces to the following Cartesian component:

\begin{align}\label{equilibriumDrude}
	0 &= -q E_x - m \frac{\langle v_x \rangle_{\rm ens}}{\tau} - \langle B_z v_y \rangle_{\rm ens}.
\end{align}
Here, the ensemble average $\langle \hat{O} \rangle_{\rm ens}$ for any operator $\hat{O}$ is defined as $
\langle \hat{O} \rangle_{\rm ens} = \big[ \langle \hat{O} \rangle \big]_{c}$, where $[\hat{O}]_{c}$ denotes a configurational average over skyrmion positions, and $
\langle \hat{O} \rangle \equiv \mathrm{Tr}(\rho\, \hat{O})$ is the quantum mechanical expectation value with respect to the conduction electron density matrix $\rho$. This formulation ensures that both the spatial distribution of skyrmions and the quantum spin state of conduction electrons are consistently taken into account when evaluating transport quantities.
In the regime where the applied electric field is strong and the influence of the emergent magnetic field is weak, the longitudinal electron dynamics is primarily governed by the electric field and the relaxation time $\tau$. Under this assumption, the term $\langle B_z v_y \rangle_{\rm ens}$ can be neglected, and the steady-state longitudinal velocity simplifies to
\begin{align}\label{vxcomponent}
	\langle v_x \rangle_{\rm ens} \approx - \frac{\tau q}{m} E_x.
\end{align}
For the transverse ($y$) direction,  Eq. \eqref{drudefirst}  reduces to  $0 = - m \frac{\langle v_y \rangle_{\rm ens}}{\tau} + \langle B_z v_x \rangle_{\rm ens}.$ Substituting Eq.~\eqref{vxcomponent} for $\langle v_x \rangle_{\rm ens}$  into Eq. \eqref{equilibriumDrude} and applying a mean-field factorization $\langle B_z v_x \rangle_{\rm ens} \approx \langle B_z \rangle_{\rm ens}\,\langle v_x \rangle_{\rm ens}$, we obtain the approximate transverse velocity:
\begin{align}
	\langle v_y \rangle_{\rm ens} 
	&\approx \frac{\tau}{m} \langle B_z \rangle_{\rm ens} \langle v_x \rangle_{\rm ens} \nonumber \\
	&= -\frac{q \tau^2}{m^2} E_x \langle B_z^{\rm av} \rangle,
\end{align}
where $B_z^{av} \equiv[B_z]_c$ denotes the configurationally averaged emergent magnetic field. The corresponding longitudinal current is $ j_x \equiv -q n \langle v_x \rangle_{\rm ens} = \frac{n \tau q^2}{m} E_x \equiv \sigma_{xx} E_x$, where $\sigma_{xx} = \frac{n \tau q^2}{m}$ is the conventional Drude longitudinal conductivity and $n$ is the carrier density. The transverse (Hall) current  $j_y \equiv -q n \langle v_y \rangle_{\rm ens}= \frac{n\,q^2\tau^2}{m^2} E_x \langle B_{z}^{\rm av}\rangle \equiv \sigma_{yx} E_x$, with the Hall conductivity given by

\begin{align}\label{sigmaxy}
	\sigma_{yx} =
	\begin{cases} 
		\dfrac{n\,q^2\tau^2}{m^2}\, B_{zz}^{\rm av}\,\langle \sigma_z \rangle, & \text{for $U_1$ (Conventional skyrmions)},  \\[8pt]
		\dfrac{n\,q^2\tau^2}{m^2}\,\big[ B_{zz}^{\rm av}\,\langle \sigma_z \rangle
		+ B_{zx}^{\rm av}\,\langle \sigma_x \rangle \big], & \text{for $U_2$ (Spin-flux skyrmions)}.
	\end{cases}
\end{align}
This result reveals that the Hall conductivity depends explicitly on the spin expectation values  of the input current and the emergent magnetic field generated by the skyrmion lattice,  leading to a spin-polarized topological Hall effect. For $U_1$ skyrmions, the Hall signal arises solely from the out-of-plane component $\langle \sigma_z \rangle$, consistent with earlier reports~\cite{Zang2011,Bruno2004,Neubauer2009,Gobel2017,Yufan2013,Nagaosa2013}. In contrast, for $U_2$ skyrmions, an additional in-plane term proportional to $\langle \sigma_x \rangle$ emerges due to the finite $B_{zx}^{\rm av}$ component.

\begin{figure}
	\begin{center}
		\subfloat[\label{hallu1}]{
			\includegraphics[scale=0.45]{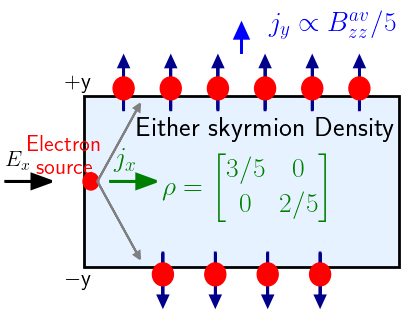}}~~
		\subfloat[\label{hallu2}]{
			\includegraphics[scale=0.45]{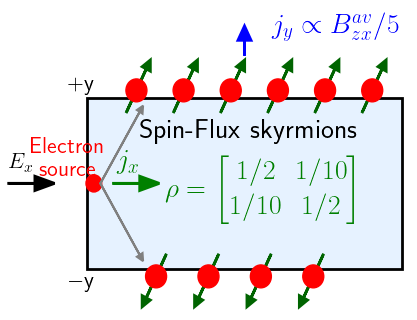}}
	\end{center}
	\caption{Cartoon illustration of the transverse Hall effect from a density of conventional  and spin-flux skyrmions. (a)For a current composed of $60\%$ electrons with $\langle \sigma_z \rangle = +1$ (upward arrows) and $40\%$ with $\langle \sigma_z \rangle = -1$ (downward arrows), electrons with $\langle \sigma_z \rangle = +1$ deflect toward the $+y$ edge, while those with $\langle \sigma_z \rangle = -1$ accumulate at the $-y$ edge. The resulting Hall current originates from the average $B_{zz}^{\mathrm{av}}$ component of the emergent field and is identical for both conventional and spin-flux skyrmions. b) For a current composed of $60\%$ electrons with $\langle \sigma_x \rangle = +1$ (oblique right arrows) and $40\%$ with $\langle \sigma_x \rangle = -1$ (oblique left arrows), the Hall response arises from the average $B_{zx}^{\mathrm{av}}$ component.  For this input current, a finite Hall current arises only from a density of spin-flux skyrmions, while a density of conventional skyrmions yields no transverse response.}
	\label{appexdix2}
\end{figure}

To illustrate this quantitatively, consider a conduction electron population described by the spin density matrix $\rho = \begin{pmatrix} 3/5 & 0\\	0 & 2/5\end{pmatrix} $, which corresponds to incoherent mixture of $60\%$ electrons with $\langle \sigma_z\rangle = +1$ (spin-up) and $40\%$ electrons with $\langle \sigma_z\rangle = -1$ (spin-down). In other words, $n_\uparrow = 0.60n$ and $n_\downarrow = 0.40n$, where $n = n_\uparrow + n_\downarrow$ is the total carrier density.
From Eq.~\ref{sigmaxy}, the emergent field $B_{zz}^{ av}$ deflects spin-up electrons toward the $+y$ direction and spin-down electrons toward the $-y$ direction with equal transverse velocity magnitude $v_y$. The resulting transverse \emph{charge} current density is then

\begin{align}
j_y = -q\big(n_\uparrow <v_{y,\uparrow}>_{ens} + n_\downarrow <v_{y,\downarrow}>_{ens}\big)
= \frac{q^2\tau^2}{m^2} B_{zz}^{\rm av} E_x \big(n_\uparrow - n_\downarrow\big)
= \frac{1}{5}\,\frac{n\,q^2\tau^2}{m^2}\, B_{zz}^{\rm av} E_x.
\end{align}
This applies to both $U_1$ and $U_2$ skyrmion densities, as illustrated in Fig.~\ref{appexdix2}(a).

Alternatively, consider a conduction electron population consisting of  an an incoherent mixture with $60\%$ of electrons polarized along $+\hat{x}$ ($\langle\sigma_x\rangle=+1$) and $40\%$ along $-\hat{x}$ ($\langle\sigma_x\rangle=-1$). The corresponding spin density matrix is $\rho = \begin{pmatrix}	1/2 & 1/10 \\1/10 & 1/2
\end{pmatrix}$, expressed in the $\sigma_z$ basis.  In the $\sigma_x$ basis, $n_{+}=0.60n$ and $n_{-}=0.40n$, where $n = n_{+}+n_{-}$ is the total carrier density. The emergent magnetic field $B_{zx}^{\rm av}$ now deflects $\langle\sigma_x\rangle = +1$ electrons in the $+y$ direction and $\langle\sigma_x\rangle = -1$ electrons in the $-y$ direction, both with equal transverse velocity magnitude $v_y$. The resulting transverse \emph{charge} current density is

\begin{align}
j_y = -q\big(n_{+} <v_{y,+}>_{ens} + n_{-} <v_{y,-}>_{ens}\big)
= \frac{q^2\tau^2}{m^2} B_{zx}^{\rm av} E_x (n_{+} - n_{-})
=\frac{1}{5}\,\frac{n\,q^2\tau^2}{m^2}\, B_{zx}^{\rm av} E_x.
\end{align}
This contribution is unique to the $U_2$ (spin-flux) skyrmions, since $B_{zx}^{\rm av}$ is absent in conventional $U_1$ skyrmions. In other words, a finite Hall response under in-plane spin polarization provides a direct experimental signature for the presence of spin-flux skyrmions, as illustrated in Fig.~\ref{appexdix2}(b).

\begin{figure}
	\begin{center}
		\subfloat[\label{exp}]{
			\includegraphics[scale=0.32]{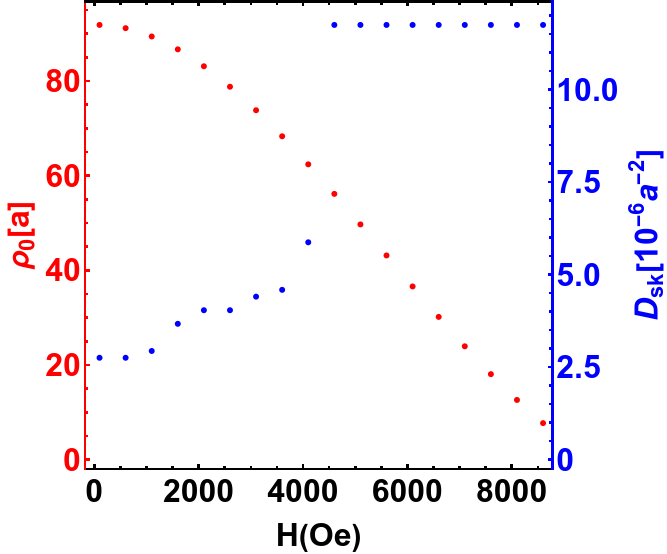}}~~
		\subfloat[\label{sigmau1}]{
			\includegraphics[scale=0.28]{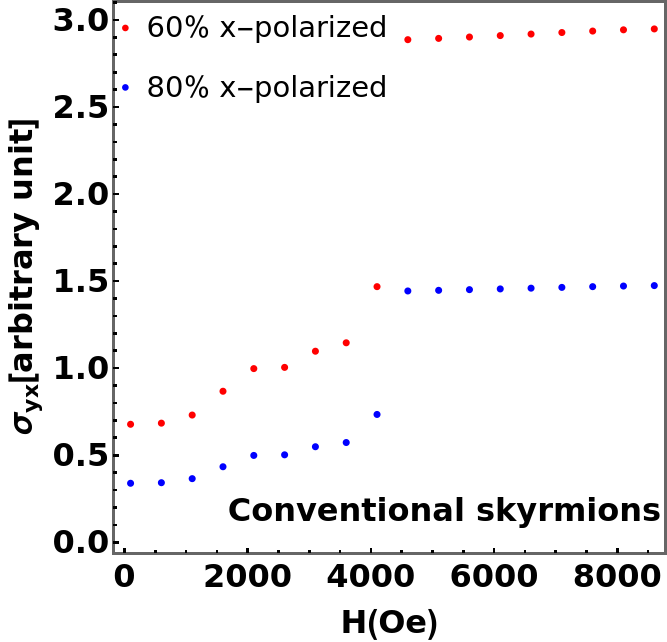}}~
		\subfloat[\label{sigmau2}]{
			\includegraphics[scale=0.28]{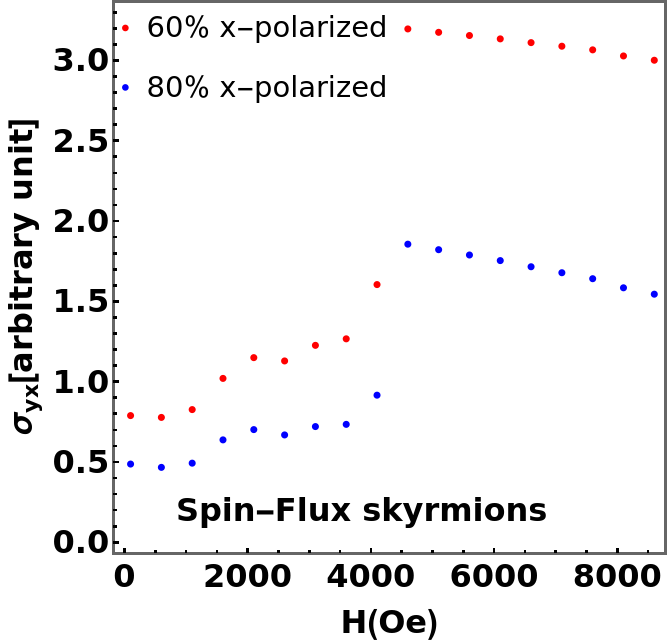}}~
	\end{center}
	\caption{Panel (a) shows  extracted experimental data from Ref.~\cite{Lone2024}, giving the skyrmion core  radius  ($\rho_0$) and the skyrmion density ($D_{sk}$) as a function of external magnetic field.  The experimental $D_{sk}$  increases from $2.5\times10^{-6}a^{-2}$ to $11.5\times10^{-6}a^{-2}$ as the field is  increased. To simulate such densities  we utilize a circular region of radius $1317a$. Using  the experimental data from (a), we compute the field dependence of the Hall conductivity  using  Eq. \ref{sigmaxy}. The configuration averaged  emergent  fields are calculated using the recipe  illustrated in Fig. \ref{bzavr}.  Panel (b): For conventional  skyrmions, the Hall conductivity increases rapidly with field up to the point where the skyrmion density saturates, and then continues to increase slowly due to the gradual reduction of the skyrmion radius. Panel (c): For spin-flux skyrmions, the  Hall conductivity first rises rapidly with field but, after the density saturates, decreases with increasing field. This decrease originates from the finite $B_{zx}$ component of the emergent field and reproduces the experimental trend reported in Ref.~\cite{Lone2024}.   In panels (b) and (c), we consider two choices of the input current. In the first case, an incoherent mixture of $40\%$ $\langle\sigma_z\rangle=1$ and $60\%$ $\langle\sigma_x\rangle=1$ is assumed (red dots), while in the second case $20\%$ $\langle\sigma_z\rangle=1$ and $80\%$ $\langle\sigma_x\rangle=1$ is used (blue dots). The slope of the conductivity decrease depends on the relative weights of in-plane and out-of-plane spin components. The $\sigma_{yx}$ scale on the $y$-axis is the same for panels (b) and (c). }
	\label{appexdix3}
\end{figure}

Experimental results are often reported in terms of the Hall resistivity $\rho_{yx}$. This is related to the conductivity  through the tensor inversion formula $\rho_{yx} = \frac{\sigma_{yx}}{\sigma_{xx}^2 + \sigma_{yx}^2}$.  In the small Hall angle limit (as also noted in Ref. \cite{Lone2024}), where $|\sigma_{yx}| \ll \sigma_{xx}$, this reduces to $\rho_{yx} \approx \frac{\sigma_{yx}}{\sigma_{xx}^2}$.  Since $\sigma_{xx}$ is  approximately constant with the external magnetic field, the field dependence of $\rho_{yx}$  closely follows that of $\sigma_{yx}$, differing only by a scaling factor.  In this case $\sigma_{yx}$  can be  compared to  experimentally observed variations of  $\rho_{yx}$  with  an externally applied magnetic  field. It has been shown  \cite{sandip2019,Romming2013} that an external magnetic field  controls the density and the core radii of the magnetic skyrmions.

In Ref.~\cite{Lone2024},  for the material  system [Ta$/$CoFeB$/$MgO]$_{1.5}$, the   measured  topological Hall conductivity increases with magnetic field, reaches a maximum, and then decreases. From the same study, we extract the field dependence of both the skyrmion core radius ($\rho_0$) and the skyrmion density ($D_{sk}$), as shown in Fig.~\ref{exp}. The experimentally observed  $D_{sk}$  increases from $2.5\times 10^{-6}a^{-2}$ to  $11.5 \times 10^{-6}a^{-2}$ (with $a=0.29\ \mathrm{nm}$) as the  external magnetic field is increased   from   $0$ to $4500$ Oe, and then $D_{sk}$ remains nearly constant over the field range $4500$-$8500$ Oe. In order to simulate this behavior, we model a circular region of radius $1317a$ to maintain a total skyrmion number and  density, $D_{sk}$, consistent with the experimentally observed values (Fig.~\ref{exp}). Using our model, we  compute the average emergent fields. For a given applied magnetic field, each pair of data points in Fig.  \ref{exp} can be associated with a single point in either  Fig.~\ref{bzzall} or \ref{bzxall}. As the applied magnetic field is varied continuously, it describes paths in   Figs.~\ref{bzzall} and \ref{bzxall}, from which the emergent magnetic fields as a function of applied magnetic field is determined.  Inserting these emergent fields into our analytical framework (Eq. \ref{sigmaxy}), we calculate the Hall conductivity for both conventional and spin-flux skyrmions, as shown in Figs.~\ref{sigmau1} and \ref{sigmau2}. This is done for two choices of the input current.   In  the first case, we consider an  incoherent mixture of $40\%$ $\langle\sigma_z\rangle=1$ and $60\%$ $\langle\sigma_x\rangle=1$, with electron spin density-matrix $\rho_1 = \begin{pmatrix} 0.7 & 0.3 \\0.3 & 0.3
\end{pmatrix}$. In the second case, an incoherent mixture of $20\%$ $\langle\sigma_z\rangle=1$ and $80\%$  $<\sigma_x>=1$, with electron spin density-matrix   $\rho_2 = \begin{pmatrix} 0.6 & 0.4 \\0.4 & 0.4
\end{pmatrix}$.

Figure~\ref{sigmau1} shows the behavior of $\sigma_{yx}$ with external magnetic field for conventional skyrmions, evaluated for the two density matrices, $\rho_1$ (red dots) and $\rho_2$ (blue dots). In both cases, $\sigma_{yx}$ increases rapidly with magnetic field, and  then continues to rise only slowly once the skyrmion density saturates. This trend arises because both the emergent field component $B_{zz}$ and the skyrmion density increase with magnetic field. After the density becomes nearly constant, $\sigma_{yx}$ is controlled primarily by the shrinking skyrmion core radius. Since conventional skyrmions only possess a finite $B_{zz}$ component, the resulting Hall  conductivity never decreases with field but instead shows a gradual saturation.

In contrast, for spin-flux skyrmions (Fig.~\ref{sigmau2}), both $B_{zz}$ and $B_{zx}$ components contribute. Initially, $\sigma_{yx}$ again increases with field due to the rising skyrmion density. However, once the density saturates, the conductivity is determined by the core radius: $B_{zz}$ decreases with  radius (see figure \ref{bzzrho} ), while $B_{zx}$ increases (see figure \ref{bzxrho}). Their combined  effect in Eq.~\ref{sigmaxy} produces a nonmonotonic $\sigma_{yx}$, which decreases after reaching a maximum. The rate of decrease depends sensitively on the relative spin composition of the conduction electrons. Specifically, a larger fraction of in-plane spins enhances the relative weight of  the $B_{zx}$ contribution,  which counterbalances  the $B_{zz}$  term and leads to faster drop in  $\sigma_{yx}$.  This effect is evident in Fig.~\ref{sigmau2}, where the blue dots denote results for $\rho_2$ and red symbols for $\rho_1$. For $\rho_2$, $\sigma_{yx}$ decreases more rapidly because of the stronger in-plane spin polarization. The resulting field dependence of $\sigma_{yx}$ in our spin-flux model resembles the experimental trend reported in Ref.~\cite{Lone2024}, where $\sigma_{yx}$ increases at low fields, peaks, and then decreases.  Such  non-monotonic behavior of  $\sigma_{yx}$  with applied magnetic field is not  explained by conventional skyrmions alone. Similar behavior of  the experimentally observed the Hall resistivity with external magnetic field  is reported in other materials,  including FeGe \cite{Huang2012,Porter2014}, MnSi \cite{Lee2009,Li2013}, MnGe \cite{Kanazawa2011}, GdRu$_2$Ge$_2$ \cite{Yoshimochi2024}, and  Gd$_2$PdSi$_3$ \cite{Kurumaji2019}.

\section{Discussion}
 We have introduced the concept of spin-flux  carrying skyrmions  as a possible additional modality for magnetic memory in spintronics.  Spin-flux has its origin in the fundamental two-valued nature of the wavefunction of  spin$-1/2$  electrons. Under a $2\pi$ rotation in  its internal coordinate system, the electron wavefunction changes sign, leading to observable physical consequences.
We have derived the explicit SU(2) gauge field for spin-flux skyrmions and demonstrated how their emergent fields and Hall response differ fundamentally from conventional skyrmions. The monopolar $\sigma_x$ component inherent to spin-flux skyrmions generates a finite average emergent field, leading to an additional contribution to the topological Hall conductivity that is absent in conventional systems. This feature provides a possible interpretation of several experimentally observed Hall resistivity behaviors that cannot be fully explained by the conventional skyrmion model.

The tunability of the Hall response with respect to spin polarization indicates that spin-flux skyrmions could be used to control charge transport in a manner not accessible with conventional textures. For instance, the Hall conductivity in such systems can be enhanced or suppressed depending on the spin alignment of itinerant electrons, offering a controllable mechanism for transverse charge flow under a longitudinal current. Such behavior has potential relevance for memory and logic architectures based on skyrmion motion, where controlling transverse deflection is essential to reduce energy dissipation and improve device stability~\cite{Fert2017, Nagaosa2013,Manchon2015,Gobel2021}.

Spin-flip scattering \cite{Jiwon2023,Singh2023} provides another probe to identify skyrmionic spin textures. In most experimental conditions, the kinetic energy of conduction electrons is significantly larger than the characteristic energy scale of the skyrmion–electron interaction. Previous theoretical works~\cite{Denisov2016,Denisov_2020} have treated this interaction as a scattering potential and analyzed the corresponding scattering amplitudes as a function of relevant system parameters.
Spin-flux  skyrmions  lead to stronger forms of   spin-flip scattering  of conduction electrons, than realized by conventional skyrmions. This is apparent from the  presence of finite off-diagonal components in the spin-flux skyrmion gauge field, absent in the conventional case.  The observation of (stronger  than conventional) spin-flip scattering may provide additional support  for the presence of spin-flux skyrmions.

Spin-flux skyrmions are a direct consequence of the doubly-connected topological structure of the group manifold  of  physical rotations in three-dimensional  space. We hope that our exposition will motivate more focused experimental effort to identify such skyrmions. Such an identification may offer previously unrecognized  technological opportunities in spintronics.

\section*{Acknowledgment}
This work was supported by the Natural  Sciences and Engineering Research  Council of Canada.

\appendix
\section{Detailed Calculation  of  emergent magnetic field }\label{emergentu1}

The emergent vector potential and the corresponding magnetic field arising from the SU(2) gauge transformation are expressed as

\begin{align}\label{vecmagfor}
	& \mathcal{\vec{A}}=- i\hbar U^{\dagger}\bm\nabla U,~~~~~ 
	\vec{ B} =\bm{\nabla} \times \mathcal{\vec{A}}
\end{align}

\subsection{Conventional Skyrmion}\label{appendixu1}

For a conventional skyrmion, the explicit form of the unitary matrix is $U_1 (\bm r) =\exp(-i\frac{\pi(\hat{n}\cdot\vec{\sigma})}{2}) = -i\hat{n}\cdot\vec{\sigma}$,
where the unit vector $\vec{n} = (\sin\frac{\theta}{2}\cos\phi, \sin\frac{\theta}{2}\sin\phi, \cos\frac{\theta}{2})$.  Here, we have used the Pauli matrix  identity $e^{i\theta\hat{n}\cdot\vec{\sigma}}= \sigma_0 \cos\theta+ i \hat{n}\cdot\vec{\sigma}\sin\theta$. Using the gradient operator  $\bm{\nabla} \equiv \hat{r} \frac{\partial}{\partial r} + \frac{\hat{\zeta}}{r} \frac{\partial }{\partial \zeta} $, the  corresponding  SU(2) vector potential can be derived using the observation that:   $\hat{n}\cdot\vec{\sigma}\partial_r (\hat{n}\cdot\vec{\sigma})= \frac{\theta_r}{2}  
\begin{pmatrix}
	0 & e^{-i \phi } \\
	-e^{i \phi } & 0 \\
\end{pmatrix} $  and $\hat{n}\cdot\vec{\sigma}\partial_\zeta (\hat{n}\cdot\vec{\sigma})=  i\phi_\zeta\begin{pmatrix}
\sin ^2\left(\frac{\theta }{2}\right) & -\frac{1}{2}  e^{-i \phi } \sin \theta  \\
-\frac{1}{2}  e^{i \phi } \sin \theta  & - \sin ^2\left(\frac{\theta }{2}\right) \\
\end{pmatrix}$. It follows that

\begin{align}
	\vec{\mathcal{A}} &=- i\hbar U_1^{\dagger}\bm\nabla U_1  \\ & \equiv\frac{\hbar}{2}\Big[ \theta_{r}(\sigma_{y}\cos\phi -\sigma_{x}\sin\phi)\hat{r}  +   \hat{\zeta}\frac{2\phi_{\zeta}\sin \frac{\theta}{2}}{r} \Big(\sigma_{z} \sin \frac{\theta}{2} -\cos \frac{\theta}{2} (\sigma_{x}\cos\phi +\sigma_{y}\sin\phi)\Big)\Big]. 
\end{align}
Consequently, the corresponding emergent magnetic field  is given by  \cite{Nagaosa2013,Fert2017}

\begin{align}
\vec{ B} & =\nabla \times \mathcal{\vec{A}} \\ &   =\frac{1}{r} \Big[\frac{\partial}{\partial r} \big( r A_\zeta \big) -  \frac{\partial A_r}{\partial \zeta} \Big] \hat{z} \\ & =\frac{\hbar}{2 r}\theta_{r}\phi_{\zeta} \Bigg[ \sigma_z \sin\theta  +2\sin^2 \frac{\theta}{2} (\sigma_x \cos\phi +\sigma_y \sin\phi)\Bigg] \hat{z} \equiv B_{z}\hat{z}
\end{align}
where  $\theta_r =\frac{\partial\theta}{\partial r}, \phi_\zeta =\frac{\partial\phi}{\partial \zeta}$. We consider a Néel-type skyrmion in our calculation, described by the solution $\theta(r)=2\tan^{-1}(r/\rho_0)$ and $ \phi=\zeta$. With this configuration, the emergent vector potential and the corresponding emergent magnetic field are given by

\begin{align}  
 \mathcal{\vec{A}}  &= \frac{\hbar}{2}\Big[ \frac{2\rho_0}{r^2 +\rho_0^2}(\sigma_{y}\cos\zeta -\sigma_{x}\sin\zeta)\hat{r}  +   \hat{\zeta} \Big(\sigma_{z}  \frac{2r}{r^2 +\rho_0^2} - \frac{2\rho_0 }{r^2 +\rho_0^2}(\sigma_{x}\cos\zeta +\sigma_{y}\sin\zeta)\Big)\Big]\\ 
   B_z &  = \hbar \Bigg[ \sigma_z \frac{2\rho_0^2}{(r^2 +\rho_0^2)^2}  +\frac{2 r\rho_0}{(r^2 +\rho_0^2)^2} (\sigma_x \cos\zeta +\sigma_y \sin\zeta)\Bigg]  \equiv \sigma_z B_{zz}  +\sigma_x B_{zx} +\sigma_y B_{zy}
\end{align}

\subsection{Spin-Flux Skyrmion }\label{appendixu2}
For a  spin-flux  skyrmion, the explicit form of the unitary matrix is $U_2 = \exp(-i\frac{\sigma_{z}\phi(\zeta)}{2}) \exp(-i\frac{\sigma_{y}\theta (r)}{2})$. 
The corresponding vector potential can be deduced using the Pauli matrix identity from \ref{appendixu1} and  the cyclic  algebra $\sigma_j \sigma_k =\delta_{jk} \sigma_0 +i \epsilon_{jkl}\sigma_l$, where  $\epsilon_{jkl}$ is  the anti-symmetric Levi-Civita  tensor.

\begin{align}
	 \vec{\mathcal{A}} &=- i\hbar U_2^{\dagger}\bm\nabla U_2  \\& =-i \hbar\Big[\frac{ -i \theta_{r}}{2}\sigma_{y}\hat{r}  +   \frac{\hat{\zeta}}{2r} (-i\phi_\zeta) U_2^{\dagger}\sigma_z U_2\Big]
\end{align}
It is straightforward to show that $U_2^{\dagger}\sigma_z U_2 =\sigma_z \cos\theta -\sigma_x \sin\theta$.

 It follows that, 

\begin{align}
\vec{\mathcal{A}}  = \frac{\hbar}{2}\Big[-\theta_{r}\sigma_{y}\hat{r}  +   \frac{\hat{\zeta  \phi_\zeta}}{r} \Big(-\sigma_{z} \cos\theta + \sigma_{x}\sin \theta\Big)\Big].
\end{align} 

Consequently, the corresponding emergent magnetic field is given by

\begin{align}
	\vec{ B} & =\nabla \times \mathcal{\vec{A}} \\ &  =\frac{1}{r} \Big[\frac{\partial}{\partial r} \big( r A_\zeta \big) -  \frac{\partial A_r}{\partial \zeta} \Big] \hat{z}  \\ &=\hbar \frac{\theta_{r}\phi_{\zeta}}{2 r} \Bigg[ \sigma_z \sin\theta  + \sigma_x \cos\theta \Bigg] \hat{z} \equiv B_{z}\hat{z}.
\end{align}
 We consider a Néel-type skyrmion in our calculation, described by the solution $\theta(r)=2\tan^{-1}(r/\rho_0)$ and $ \phi=\zeta$. With this configuration, the emergent vector potential and the corresponding emergent magnetic field are given by

\begin{align}  
	 \mathcal{\vec{A}} &  = \frac{\hbar}{2}\Big[- \frac{2\rho_0}{\rho_0^2 +r^2}\sigma_{y}\hat{r}  +   \hat{\zeta} \Big(-\sigma_{z} \frac{\rho_0^2 -r^2}{r(\rho_0^2 +r^2)} + \sigma_{x}\frac{2\rho_0}{\rho_0^2 +r^2}\Big)\Big] \equiv \mathcal{A}_r\hat{r}+ \mathcal{A}_\zeta\hat{\zeta}\\  
	B_z  & =  \hbar \Bigg[ \sigma_z \frac{2\rho_0^2}{(\rho_0^2 +r^2)^2}  +\frac{\rho_0(\rho_0^2-r^2)}{r(\rho_0^2 +r^2)^2}\sigma_x \Bigg] \equiv \sigma_z B_{zz}+ \sigma_x B_{zx}
\end{align} 
The radial component $\mathcal{A}_r$ contains a $\sigma_y$ term and is independent of the azimuthal angle $\zeta$, i.e., $
\frac{\partial \mathcal{A}_r}{\partial \zeta} = 0.$ 
Therefore, the $\sigma_y$ component does not contribute to the emergent magnetic field.

\section{ Derivation of the Spin-dependent Lorentz Equation in a Homogeneous Matrix  Field}\label{homogeneousderivation}

The Hamiltonian of an electron in the presence of an emergent field is
\begin{align}
	H' = \frac{1}{2m} \left( \bm{P} + \bm{\mathcal{A}} \right)^2 - J_h \sigma_z,
\end{align}
where  $\bm{\mathcal{A}}$ is the microscopic synthetic gauge field due to skyrmions. As described earlier, we replace  $H'$ with a mean-field Hamiltonian, $H_{MF}$ in  which  $\bm{\mathcal{A}}$ is replaced by its configuration averaged value $\bm{\mathcal{A}}_{av}= [\bm{\mathcal{A}}]_c$. In the  symmetric gauge this becomes: 

\begin{align}
\bm{\mathcal{A}}_{\text{av}} = \frac{1}{2} B_z^{\text{av}} (\vec{r}\times \hat{z}),
\end{align}
where the total averaged emergent magnetic field is given by $
B_z^{\text{av}} = B_{zx}^{\text{av}} \sigma_x + B_{zy}^{\text{av}} \sigma_y + B_{zz}^{\text{av}} \sigma_z$. 
For the \(U_1\) skyrmion, we have \( B_{zx}^{\text{av}} = B_{zy}^{\text{av}} = 0 \), while for the \(U_2\) spin-flux skyrmion, \( B_{zy}^{\text{av}} = 0 \), but \( B_{zx}^{\text{av}} \neq 0 \). The mean-field Hamiltonian is then given by:
\begin{align}
	H_{MF} &= \frac{1}{2m} \left( \bm{P} + \bm{\mathcal{A}}_{\text{av}} \right)^2 - J_h \sigma_z \\
	&= \frac{1}{2m} \left[ \sigma_0 \left( P_x^2 + P_y^2 + \frac{(x^2 + y^2)}{4} (B_z^{\text{av}})^2 \right) + B_z^{\text{av}} (x P_y - y P_x) - J_{1h} \sigma_z \right], \quad \text{where } J_{1h} = 2m J_h \\
	&\equiv \frac{1}{2m} \left[ \sigma_0 \left( \bm{P}^2 + \frac{r^2}{4} (B_z^{\text{av}})^2 \right) + B_z^{\text{av}} L_z - J_{1h} \sigma_z \right],
\end{align}
where \( L_z = x P_y - y P_x \) is the \(z\)-component of the orbital angular momentum operator.	The Heisenberg  position and momentum operators satisfy  a matrix version of the classical equations of motion \cite{Chudnovsky2007,Yamane2019}:

\begin{align}
	& \label{1} \frac{dx}{dt}=\frac{i}{\hbar}[H_{MF}, x]  =\frac{1}{2m}\Big(2 P_x\sigma_{0}- B_z^{\text{av}}y\Big) \implies  P_x \sigma_{0}=m\frac{dx}{dt}+ B_z^{\text{av}}y/2 
	\\& \label{2} \frac{dy}{dt}=\frac{i}{\hbar}[H_{MF}, y] =\frac{1}{2m}\Big(2 P_y \sigma_0+ B_z^{\text{av}}x\Big)  \implies  P_y\sigma_0=m	\frac{dy}{dt}- B_z^{\text{av}}x/2 
	\\ & \label{3} \frac{dP_x}{dt}=\frac{i}{\hbar}[ H_{MF}, P_x] =-\frac{1}{2m}\Big(\frac{x(B_z^{av})^2}{2} + B_z^{\text{av}}P_y\Big) \\ & \label{4} \frac{dP_y}{dt}=\frac{i}{\hbar}[ H_{MF}, P_y]  =-\frac{1}{2m}\Big(\frac{y(B_z^{av})^2}{2} - B_z^{\text{av}}P_x\Big).
\end{align}	
Substituting  equations  \ref{1} and \ref{2}  into equations \ref{3} and \ref{4}, we obtain:

\begin{align}	
&\frac{dP_x}{dt} = -\frac{1}{2} B_z^{\text{av}}\frac{dy}{dt}\\&	
 \frac{dP_y}{dt}=\frac{1}{2} B_z^{\text{av}}\frac{dx}{dt}
\end{align}
Since $\vec{ P}=m\dot{\vec{r}}$, we obtain the matrix form of the Lorentz force equation

\begin{align}
	m  \ddot{\vec{r}} =- B_z^{av}(\vec {v}\times \hat{z}).
\end{align}

\bibliographystyle{unsrt}
\bibliography{ge_ref_copy}
\end{document}